\documentclass[
 reprint,
 amssymb, amsmath,
 aip,jcp
]{revtex4-1}

\usepackage{comment} 
\usepackage{dcolumn} 
\usepackage[colorlinks=true,linkcolor=blue]{hyperref}%
\expandafter\ifx\csname package@font\endcsname\relax\else
 \expandafter\expandafter
 \expandafter\usepackage
 \expandafter\expandafter
 \expandafter{\csname package@font\endcsname}%
\fi
\usepackage{xspace}

\usepackage{graphicx}
\usepackage{amsmath}

\usepackage[normalem]{ulem}
\usepackage[utf8]{inputenc}

\newcommand{\norm}[1]{\left| #1\right|}

\newcommand{\mc}[1]{\multicolumn{1}{c}{#1}}

\newcommand{\ket}[1]{{\ensuremath{|#1\rangle}\xspace}}
\newcommand{\bra}[1]{{\ensuremath{\langle #1|}\xspace}}

\newcommand{\elemm}[3]{{\ensuremath{\bra{#1}{#2}\ket{#3}}\xspace}}
\newcommand{\ovrlp}[2]{{\ensuremath{\langle #1|#2\rangle}\xspace}}

\newcommand{\basis}[0]{\mathcal{B}}
\newcommand{\den}[0]{{n}}
\newcommand{\denr}[0]{{n}({\bf r})}
\newcommand{\denzr}[0]{{n_0}({\bf r})}

\newcommand{\denrbasis}[0]{{n}^{\basis}({\bf r})}
\newcommand{\denzrbasis}[0]{{n_0^\basis}({\bf r})}

\newcommand{\denfci}[0]{\den_{\psifci}}
\newcommand{\dencipsi}[0]{{n}_{\text{CIPSI}}^\basis({\bf r})}

\newcommand{\psibasis}[0]{\Psi^{\basis}}
\newcommand{\psimu}[0]{\Psi^{\mu}}
\newcommand{\psifci}[0]{\Psi^{\basis}_{\text{FCI}}}
\newcommand{\kinop}[0]{\hat{T}}
\newcommand{\kinopbasis}[0]{\hat{T}^\basis}

\newcommand{\vnebasis}[0]{\hat{V}_{\text{ne}}^\basis}

\newcommand{\weeop}[0]{\hat{W}_{\text{ee}}}
\newcommand{\weeopbasis}[0]{\hat{W}_{\text{ee}}^\basis}
\newcommand{\weeopmu}[0]{\hat{W}_{\text{ee}}^{\text{lr},\mu}}

\newcommand{\efci}[0]{E_{\text{FCI}}^{\basis}}

\newcommand{\ecmubis}[0]{\bar{E}_{\text{c,md}}^{\text{sr},\mu}[\denr]}
\newcommand{\efuncbasis}[0]{\bar{E}^\basis[\denrbasis]}

\newcommand{\efuncbasislda}[0]{\bar{E}_{\text{LDA}}^{\basis,\psibasis}[\denrbasis]}
\newcommand{\efuncbasisldafci}[0]{\bar{E}_{\text{LDA}}^{\basis,\psibasis}[\denfci]}
\newcommand{\efuncbasisldacipsi}[0]{\bar{E}_{\text{LDA}}^{\basis,\psibasis}[\dencipsi]}
\newcommand{\efuncbasisldacipsihf}[0]{\bar{E}_{\text{LDA}}^{\basis,\text{HF}}[\dencipsi]}
\newcommand{\efuncbasisldapsi}[1]{\bar{E}_{\text{LDA}}^{\basis,\text{#1}}[\denr]}

\newcommand{\emulda}[0]{\bar{\varepsilon}^{\text{sr},\text{unif}}_{\text{c,md}}\left(\denrbasis;\mu({\bf r};\psibasis)\right)}
\newcommand{\emuldahf}[0]{\bar{\varepsilon}^{\text{sr},\text{unif}}_{\text{c,md}}\left(\dencipsi;\mu({\bf r};\text{HF})\right)}
\newcommand{\emuldafinal}[0]{E^{\basis,\psibasis}_{\text{FCI+LDA}}}

\newcommand{\real}[1]{{\rm I\!R}^{#1}}
\newcommand{\bfr}[1]{{\bf X}_{#1}}
\newcommand{\bfrb}[1]{{\bf r}_{#1}}
\newcommand{\rab}[1]{|\bfrb{1} - \bfrb{2}|}

\newcommand{\dr}[1]{\text{d}\bfr{#1}}
\newcommand{\rr}[2]{\bfr{#1}, \bfr{#2}}
\newcommand{\rrrr}[4]{\bfr{#1}, \bfr{#2},\bfr{#3},\bfr{#4}  }

\newcommand{\psix}[1]{\hat{\Psi}\left({\bf X}_{#1}\right)}
\newcommand{\psixc}[1]{\hat{\Psi}^{\dagger}\left({\bf X}_{#1}\right)}
\newcommand{\ai}[1]{\hat{a}_{#1}}
\newcommand{\aic}[1]{\hat{a}^{\dagger}_{#1}}
\newcommand{\vijkl}[0]{V_{ij}^{kl}}
\newcommand{\phix}[2]{\phi_{#1}(\bfr{#2})}
\newcommand{\phixprim}[2]{\phi_{#1}(\bfr{#2}')}

\newcommand{\twodm}[4]{\elemm{\Psi}{\psixc{#4}\psixc{#3} \psix{#2}\psix{#1}}{\Psi}}
\newcommand{\twodmr}[5]{ n^{(2)}_{#5}(\rrrr{#1}{#2}{#3}{#4})}
\newcommand{\twodmrdiag}[3]{ n^{(2)}_{#3}(\rr{#1}{#2})}
\newcommand{\ontop}[2]{ n^{(2)}_{#1}({\bf #2}_1)}

\newcommand{\twodmrdiagpsi}[0]{ n^{(2)}_{\psibasis}(\rr{1}{2})}
\newcommand{\gammamnpq}[1]{\Gamma_{mn}^{pq}[#1]}

\newcommand{\gammaklmn}[1]{\Gamma_{kl}^{mn}[#1]}
\newcommand{\fbasis}[0]{f_{\psibasis}(\bfr{1},\bfr{2})}
\newcommand{\wbasis}[0]{W_{\psibasis}(\bfr{1},\bfr{2})}
\newcommand{\wbasisr}[0]{W_{\psibasis}(\bfrb{1},\bfrb{2})}

\newcommand{\wbasispsi}[1]{W_{#1}(\bfr{1},\bfr{2})}
\newcommand{\wbasismu}[0]{W_{\psibasis}^{\text{lr},\mu({\bf r}_1)}(\bfrb{1},\bfrb{2})}
\newcommand{\wbasishf}[0]{W_{\text{HF}^{\basis}}(\bfrb{1},\bfrb{2})}
\newcommand{\wbasisfci}[0]{W_{\text{FCI}^{\basis}}(\bfrb{1},\bfrb{2})}
\newcommand{\wbasiscoal}[1]{W_{\psibasis}({\bf r}_{#1})}

\begin{document}

\title{Curing basis-set convergence of wave-function theory using density-functional theory: a systematically improvable approach} 

\date{August 23, 2018}
\begin{abstract}
The present work proposes to use density-functional theory (DFT) to correct for the basis-set error of wave-function theory (WFT). One of the key ideas developed here is to define a range-separation parameter which automatically adapts to a given basis set. The derivation of the exact equations are based on the Levy-Lieb formulation of DFT, which helps us to define a complementary functional which corrects uniquely for the basis-set error of WFT. The coupling of DFT and WFT is done through the definition of a real-space representation of the electron-electron Coulomb operator projected in a one-particle basis set. Such an effective interaction has the particularity to coincide with the exact electron-electron interaction in the limit of a complete basis set, and to be finite at the electron-electron coalescence point when the basis set is incomplete. The non-diverging character of the effective interaction allows one to define a mapping with the long-range interaction used in the context of range-separated DFT and to design practical approximations for the unknown complementary functional. Here, a local-density approximation is proposed for both full-configuration-interaction (FCI) and selected configuration-interaction approaches. Our theory is numerically tested to compute total energies and ionization potentials for a series of atomic systems. The results clearly show that the DFT correction drastically improves the basis-set convergence of both the total energies and the energy differences. For instance, a sub kcal/mol accuracy is obtained from the aug-cc-pVTZ basis set with the method proposed here when an aug-cc-pV5Z basis set barely reaches such a level of accuracy at the near FCI level. 
\end{abstract}

\author{Emmanuel Giner}%
\email{emmanuel.giner@lct.jussieu.fr}
\affiliation{Laboratoire de Chimie Théorique, Sorbonne Université and CNRS, F-75005 Paris, France}
\author{Barth\' elemy Pradines}
\affiliation{Laboratoire de Chimie Théorique, Sorbonne Université and CNRS, F-75005 Paris, France}
\affiliation{Institut des Sciences du Calcul et des Données, Sorbonne Université, F-75005 Paris, France}
\author{Anthony Fert\' e}
\affiliation{Laboratoire de Chimie Théorique, Sorbonne Université and CNRS, F-75005 Paris, France}
\author{Roland Assaraf}
\affiliation{Laboratoire de Chimie Théorique, Sorbonne Université and CNRS, F-75005 Paris, France}
\author{Andreas Savin}
\affiliation{Laboratoire de Chimie Théorique, Sorbonne Université and CNRS, F-75005 Paris, France}
\author{Julien Toulouse}
\affiliation{Laboratoire de Chimie Théorique, Sorbonne Université and CNRS, F-75005 Paris, France}
\maketitle

\section{Introduction}
The development of accurate and systematically improvable computational methods to calculate the electronic structure of molecular systems is an important research topic in theoretical chemistry as no definitive answer has been brought to that problem. The main difficulty originates from the electron-electron interaction which induces correlation between electrons, giving rise to a complexity growing exponentially with the size of the system. In this context, the two most popular approaches used nowadays, namely wave-function theory (WFT) and density-functional theory (DFT), have different advantages and limitations due to the very different mathematical formalisms they use to describe the electronic structure. 

The clear advantage of WFT relies on the fact that, in a given one-electron basis set, the target accuracy is uniquely defined by the full-configuration-interaction (FCI) limit. Therefore, there exists many ways of systematically improving the accuracy by refining the wave-function ansatz, and ultimately by enlarging the basis set. In particular, perturbation theory is a precious guide for approximating the FCI wave function and it has given birth to important theorems\cite{goldstone,lindgren} and many robust methods, such as coupled cluster\cite{review_cc_bartlett} or selected configuration interaction (CI)\cite{bender,malrieu,buenker1,buenker-book,three_class_CIPSI,harrison,hbci}. Despite these appealing features, the main disadvantages of WFT are certainly the slow convergence of many important physical properties with respect to the size of the one-particle basis set and the rapidly growing computational cost when one enlarges the basis set. Such behavior very often prohibits the reach of the so-called complete-basis-set limit which is often needed to obtain quantitative agreement with experiment. At the heart of the problem of slow convergence with respect to the size of the basis set lies the description of correlation effects when electrons are close, the so-called short-range correlation effects near the electron-electron cusp\cite{kato}. To cure this problem, explicitly correlated ($f_{12}$) methods have emerged from the pioneering work of Hylleraas\cite{hylleraas} and remain an active and promising field of research (for recent reviews, see Refs.~\onlinecite{rev_f12_tew,rev_f12_vallev,rev_f12_gruneis}). One possible drawback of the $f_{12}$ methods is the use of a rather complex mathematical machinery together with numerically expensive quantities involving more than two-electron integrals. 

An alternative formulation of the quantum many-body problem is given by DFT which, thanks to the Hohenberg-Kohn theorems\cite{hk_theorem}, abandons the complex many-body wave function for the simple one-body density. Thanks to the so-called Kohn-Sham formalism of DFT\cite{ks_dft} and the development of practical approximations of the exchange-correlation density functional, DFT is nowadays the most used computational tool for the study of the molecular electronic problem. Despite its tremendous success in many areas of chemistry, Kohn-Sham DFT applied with usual semilocal density functional approximations generally fails to describe nonlocal correlation effects, such as strong correlation or dispersion forces. To overcome these problems ingredients from WFT have been introduced in DFT, starting from Hartree-Fock (HF) exchange\cite{Bec-JCP-93a} to many-body perturbation theory\cite{GoeGri-WIRE-14}. Nevertheless, the lack of a scheme to rationally and systematically improve the quality of approximate density functionals\cite{gunnarson_dft_rev} remains a major limitation of DFT. 

A more general formulation of DFT has emerged with the introduction of the so-called range-separated DFT (RS-DFT) (see Ref.~\onlinecite{rs_dft_toul_colo_savin} and references therein) which rigorously combines WFT and DFT. In such a formalism the electron-electron interaction is split into a long-range part which is treated using WFT and a complementary short-range part treated with DFT. The formalism is exact provided that full flexibility is given to the long-range wave function and that the exact short-range density functional is known. In practice, approximations must be used for these quantities and the splitting of the interaction has some appealing features in that regard. As the long-range wave-function part only deals with a non-diverging electron-electron interaction, the problematic cusp condition is removed and the convergence with respect to the one-particle basis set is greatly improved\cite{basis_set_rs_dft}. Regarding the DFT part, the approximate semilocal density functionals are better suited to describe short-range interaction effects. Therefore, a number of approximate RS-DFT schemes have been developed using either single-reference WFT approaches (such as M{\o}ller-Plesset perturbation theory\cite{AngGerSavTou-PRA-05}, coupled cluster\cite{GolWerSto-PCCP-05}, random-phase approximations\cite{TouGerJanSavAng-PRL-09,JanHenScu-JCP-09}) or multi-reference WFT approaches (such as multi-reference CI\cite{LeiStoWerSav-CPL-97}, multiconfiguration self-consistent field\cite{FroTouJen-JCP-07}, multi-reference perturbation theory\cite{fromager_rs_nevpt2}, density-matrix renormalization group\cite{dmrg_rs_dft_1}). These mixed WFT/DFT schemes have shown to be able to correctly describe a quite wide spectrum of chemical situations going from weak intermolecular interactions to strong correlation effects. Nonetheless, these methods involve a range-separation parameter, often denoted by $\mu$, and there is no fully satisfying and systematic scheme to set its value, even if some interesting proposals have been made\cite{optimal_tuned_rsdft,local_mu_hybrid_1,local_mu_hybrid_2}. 

The main goal of the present work is to use a DFT approach to correct for the basis-set incompleteness of WFT. The key idea developed here is to make a separation of the electron-electron interaction directly based on the one-particle basis set used and to express the remaining effects as a functional of the density. In practice, we propose a fit of the projected electron-electron interaction by a long-range interaction, leading to a local range-separation parameter $\mu({\bf r})$ which automatically adapts to the basis set. 
This is done by comparing at coalescence a real-space representation of the Coulomb electron-electron operator projected in the basis set with the long-range interaction used in RS-DFT. Thanks to this link, the theory proposed here can benefit from pre-existing short-range density functionals developed in RS-DFT. 

The present paper is composed as follows. We present the general equations related to the splitting of the electron-electron interaction in a one-particle basis set in sections \ref{sec:operator_basis} and \ref{sec:fci_density}. In section \ref{sec:qualitative} we point out the similarities and differences of this formalism with RS-DFT. A real-space representation of the electron-electron Coulomb operator developed in a one-particle basis set is proposed in section \ref{sec:real_Coulomb} (with details given in Appendix \ref{sec:appendix_expectation} and \ref{sec:complete_basis}), which leads to the definition of a local range-separation parameter $\mu({\bf r})$ that automatically adapts to the basis set. This allows us to define in section \ref{sec:practical_approx} a short-range local-density approximation (LDA) correcting FCI energies for the basis-set error. The formalism is then extended to the selected CI framework in section \ref{sec:cipsi_lda}. In section \ref{sec:numerical} we test our theory on a series of atomic systems by computing both total energies and energy differences. We study the basis-set convergence of the DFT-corrected FCI total energy in the case of the helium atom in section \ref{sec:fci_numerical}. We then investigate the basis-set convergence of DFT-corrected selected CI for both total energies and ionization potentials (IPs) of the B-Ne series in section \ref{sec:cipsi_numerical}. In the case of the IPs, we show that chemical accuracy is systematically reached for all atomic systems already from the aug-cc-pVTZ basis set within our approach, whereas an aug-cc-pV5Z basis set is needed to reach such an accuracy at near FCI level. In order to better understand how the DFT-based correction acts for both total energies and energy differences, a detailed study is performed in section~\ref{sec:numerical_o} for the oxygen atom and its first cation. Finally, we summarize the main results and conclude in section \ref{sec:conclusion}. 

\section{Theory}
\label{sec:theory}
\subsection{Finite basis-set decomposition of the universal density functional}
\label{sec:operator_basis}
We begin by the standard DFT formalism for expressing the exact ground-state energy:
\begin{equation}
 \label{eq:E_lieb}
 E_{0} = \min_{\denr} \Big\{  F[\denr] + (v_{\text{ne}}({\bf r})|\denr)\Big\} ,
\end{equation}
where
\begin{equation}
  (v_{\text{ne}}({\bf r})|\denr) = \int \text{d}{\bf r} \,\,v_{\text{ne}}({\bf r}) \,\, \denr
\end{equation}
is the nuclei-electron interaction energy, and $F[\denr]$ is the Levy-Lieb universal density functional 
\begin{equation}
 \label{eq:F_lieb}
 F[\denr] = \min_{\Psi \rightarrow \denr} \elemm{\Psi}{\kinop + \weeop}{\Psi},
\end{equation}
where the minimization is over $N$-electron wave functions $\Psi$ with density equal to $\denr$, and $\kinop$ and $\weeop$ are the kinetic-energy and Coulomb electron-electron interaction operators, respectively. The Levy-Lieb universal functional only depends on the density $\denr$, meaning that, given a density $\denr$, one does not in principle needs to pass through the minimization over explicit $N$-electron wave functions $\Psi$ to obtain the value $F[\denr]$. Provided that the search in equation \eqref{eq:E_lieb} is done over $N$-representable densities expanded in a complete basis set, the minimizing density will be the exact ground-state density $\denzr$, leading to the exact ground-state energy $E_0$.

First, we consider the restriction on the densities over which we perform the minimization to those that can be represented within a one-electron basis set $\basis$, which we denote by $\denrbasis$. By this we mean all the densities that can be obtained from any wave function $\psibasis$ expanded into $N$-electron Slater determinants constructed from orbitals expanded on the basis $\basis$. Note that this is a sufficient but not necessary condition for characterizing these densities, as these densities can in general also be obtained from wave functions not restricted to the basis set. Therefore, the restriction on densities representable by a basis $\basis$ is much weaker than the restriction on wave functions representable by the same basis $\basis$.
With this restriction, there is a density, referred to as $\denzrbasis$, which minimizes the energy functional of Eq.~(\ref{eq:E_lieb}) and give a ground-state energy $E_0^{\basis}$: 
\begin{equation}
 \label{eq:E_b}
 \begin{aligned}
  E_0^{\basis}& = \min_{\denrbasis} \Big\{  F[\denrbasis] + (v_{\text{ne}}({\bf r})|\denrbasis)\Big\} \\
              & = F[\denzrbasis] + (v_{\text{ne}}({\bf r})|\denzrbasis).
 \end{aligned}
\end{equation}
Therefore, provided only that the exact ground-state density $\denzr$ is well approximated by this density $\denzrbasis$,
\begin{equation}
\label{eq:n0approx}
 \denzr \approx \denzrbasis,
\end{equation}
the exact ground-state energy $E_0$ will be well approximated by $E_0^\basis$,
\begin{equation}
\label{eq:E0approx}
E_0 \approx E_0^\basis.
\end{equation}
Considering the fast convergence of the density with the size of the basis set, we expect the approximation of equation \eqref{eq:E0approx} to be very good in practice for the basis sets commonly used.

Next, we consider the following decomposition of the Levy-Lieb density functional for a given density $\denrbasis$:
\begin{equation}
 F[\denrbasis] = \min_{\psibasis \rightarrow \denrbasis}  \elemm{\psibasis}{\kinop + \weeop}{\psibasis} +  \efuncbasis,
\end{equation}
where $\psibasis$ are wave functions restricted to the $N$-electron Hilbert space generated by the basis $\basis$, and $\efuncbasis$ is a complementary density functional
\begin{equation}
  \begin{aligned}
  \label{eq:E_funcbasis}
    \efuncbasis = & \min_{\Psi \rightarrow \denrbasis}  \elemm{\Psi}{\kinop + \weeop}{\Psi} \\&- \min_{\psibasis  \rightarrow \denrbasis}  \elemm{\psibasis}{\kinop + \weeop}{\psibasis}.
  \end{aligned}
\end{equation}
It should be pointed out that, in contrast with the density functionals used in DFT or RS-DFT, the complementary functional $\efuncbasis$ is not universal as it depends on the basis set $\basis$ used to describe a specific system. 
As the restriction to the basis set $\basis$ is in general much more stringent for the $N$-electron wave functions $\psibasis$ than for the densities $\denrbasis$, we expect that the complementary functional $\efuncbasis$ gives a substantial contribution, even for basis sets $\basis$ for which the approximation of equation \eqref{eq:n0approx} is good.

By using such a decomposition in equation \eqref{eq:E_b}, we obtain now
\begin{equation}
 \label{eq:E_nb}
  \begin{aligned}
 E_0^{\basis} = \min_{\denrbasis} \Big\{&  \min_{\psibasis \rightarrow \denrbasis}  \elemm{\psibasis}{\kinop + \weeop}{\psibasis} \\
&+   (v_{\text{ne}}({\bf r})|\denrbasis) + \efuncbasis \Big\},
  \end{aligned}
\end{equation}
or, after recombining the two minimizations,
\begin{equation}
 \label{eq:E_minpsi}
  \begin{aligned}
 E_0^{\basis} = \min_{\psibasis} \Big\{& \elemm{\psibasis}{\kinop + \weeop}{\psibasis} + (v_{\text{ne}}({\bf r})|n_{\psibasis}({\bf r})) \\
&+  \bar{E}^\basis[n_{\psibasis}({\bf r})]\Big\},
  \end{aligned}
\end{equation}
where $n_{\psibasis}({\bf r})$ is the density of $\psibasis$. By writing the Euler-Lagrange equation associated to the minimization in equation \eqref{eq:E_minpsi}, we find that the minimizing wave function $\Psi_0^\basis$ satisfies the Schr\"odinger-like equation
\begin{equation}
 \label{eq:HeffB}
  \begin{aligned}
 \left( \hat{T}^\basis + \weeop^\basis + \hat{V}_\text{ne}^\basis + \hat{\bar{V}}^\basis[n_{\Psi_0^\basis}({\bf r})] \right) \ket{\Psi_0^\basis} = {\cal E}_0^\basis \ket{\Psi_0^\basis},
  \end{aligned}
\end{equation}
where $\hat{T}^\basis$, $\weeop^\basis$, $\hat{V}_\text{ne}^\basis$, and $\hat{\bar{V}}^\basis[\denr]$ are the restrictions to the space generated by the basis $\basis$ of the operators $\hat{T}$, $\weeop$, $\int \text{d}{\bf r} \, v_\text{ne}({\bf r}) \hat{n}({\bf r})$, and $\int \text{d}{\bf r} (\delta \bar{E}^\basis[\denr]/\delta \denr ) \hat{n}({\bf r})$, respectively, and $\hat{n}({\bf r})$ is the density operator. The potential $\hat{\bar{V}}^\basis[n_{\Psi_0^\basis}({\bf r})]$ ensures that the minimizing wave function $\Psi_0^\basis$ gives the minimizing density $\denzrbasis$ in equation \eqref{eq:E_b}. It is important to notice that the accuracy of the obtained energy $E_0^{\basis}$ depends only on how close the density of $\Psi_0^\basis$ is from the exact density: $n_{\Psi_0^\basis}({\bf r})=n_0({\bf r}) \Longrightarrow E_0^\basis =E_0$.

\subsection{Approximation of the FCI density in a finite basis set}
\label{sec:fci_density}

In the limit where $\basis$ is a complete basis set, equation~\eqref{eq:E_minpsi} gives the exact energy and $\efuncbasis= 0$.
When the basis set is not complete but sufficiently good, $\efuncbasis$ can be considered as a small perturbation.
Minimizing in equation~\eqref{eq:E_minpsi} without $\efuncbasis$ simply gives the FCI energy in a given basis set $\basis$
\begin{equation}
 \label{eq:E_FCI}
  \begin{aligned}
 \efci &= \min_{\psibasis} \Big\{ \elemm{\psibasis}{\kinop + \weeop}{\psibasis} + (v_{\text{ne}}({\bf r})|n_{\psibasis}({\bf r})) \Big\}\\
       &=\elemm{\psifci}{\kinop + \weeop}{\psifci} + (v_{\text{ne}}({\bf r})|n_{\psifci}({\bf r})),
  \end{aligned}
\end{equation}
where we have introduced the  ground-state FCI wave function $\psifci$ which satisfies the eigenvalue equation:
\begin{equation}
 \label{eq:H_FCI}
  \left( \kinopbasis + \weeopbasis + \vnebasis \right) \,\,\ket{\psifci} = \efci \,\,\ket{\psifci}.
\end{equation}
Note that the  FCI energy $\efci$  is an upper bound of $E_0^{\basis}$ in equation \eqref{eq:E_minpsi} since $\efuncbasis\leq 0$.
By neglecting the impact of $\hat{\bar{V}}^\basis[n_{\Psi_0^\basis}({\bf r})]$ on the minimizing density  $\denzrbasis$, we propose a zeroth-order approximation for the density
\begin{equation}
\label{eq:Psi0bapprox}
n_0^\basis({\bf r}) \approx n_{\psifci}({\bf r}),
\end{equation}
which leads to a first-order-like approximation for the energy $E_0^{\basis}$
\begin{equation}
 \label{eq:E_B_fci}
  E_{0}^{\basis}  \approx   \efci +  \bar{E}^\basis[n_{\psifci}({\bf r})].
\end{equation}
The term $\bar{E}^\basis[n_{\psifci}({\bf r})]$ constitutes a simple DFT correction to the FCI energy which should compensate for the incompleteness of the basis set $\basis$. The next sections are devoted to the analysis of the properties of $\efuncbasis$ and to some practical approximations for this functional.

\subsection{Qualitative considerations for the complementary functional $\efuncbasis$}
\label{sec:qualitative}
The definition of $\efuncbasis$ [see equation \eqref{eq:E_funcbasis}] is clear but deriving an approximation for such a functional is not straightforward. For example, defining an LDA-like approximation is not easy as the wave functions $\psibasis$ used in the definition of $\efuncbasis$ are not able to reproduce a uniform density if the basis set $\basis$ is not translationally invariant. 
Nonetheless, it is known that a finite one-electron basis set $\basis$ usually describes poorly the short-range correlation effects and therefore the functional $\efuncbasis$ must recover these effects. Therefore, a natural idea is to find a mapping between this functional with the short-range functionals used in RS-DFT.
Among these functionals, the multi-determinant short-range correlation functional $\ecmubis$ of Toulouse \textit{et al.}\cite{Toulouse2005_ecmd} has a definition very similar to the one of $\efuncbasis$: 
\begin{equation}
 \begin{aligned}
  \label{eq:ec_md_mu}
  \ecmubis = & \min_{\Psi \rightarrow \denr}  \elemm{\Psi}{\kinop + \weeop}{\Psi} \\  - \; & \elemm{\psimu[\denr]}{\kinop + \weeop}{\psimu[\denr]},
 \end{aligned}
\end{equation}
where the wave function $\psimu[\denr]$ is defined by the constrained minimization
\begin{equation}
\label{eq:argmin}
\psimu[\denr] = \arg \min_{\Psi \rightarrow \denr} \elemm{\Psi}{\kinop + \weeopmu}{\Psi},
\end{equation}
where $\weeopmu$ is the long-range electron-electron interaction operator
\begin{equation}
  \label{eq:weemu}
  \weeopmu = \frac{1}{2} \iint \text{d}{\bf r}_1  \text{d}{\bf r}_2 \; w^{\text{lr},\mu}(|{\bf r}_1 - {\bf r}_2|) \hat{n}^{(2)}({\bf r}_1,{\bf r}_2),
\end{equation}
with
\begin{equation}
\label{eq:erf}
w^{\text{lr},\mu}(|{\bf r}_1 - {\bf r}_2|) = \frac{\text{erf}(\mu |{\bf r}_1 - {\bf r}_2|)}{|{\bf r}_1 - {\bf r}_2|},
\end{equation}
and the pair-density operator $\hat{n}^{(2)}({\bf r}_1,{\bf r}_2) =\hat{n}({\bf r}_1) \hat{n}({\bf r}_2) - \delta ({\bf r}_1-{\bf r}_2) \hat{n}({\bf r}_1)$.
By comparing equation \eqref{eq:ec_md_mu} to the definition of $\efuncbasis$ in equation \eqref{eq:E_funcbasis}, one can see that the only difference between these two functionals relies in the wave functions used for the constrained minimization: in $\ecmubis$ one uses $\psimu$ whereas $\psibasis$ is used in $\efuncbasis$. 
More specifically, $\psimu$ is determined by using a non-diverging long-range electron-electron interaction defined in a complete basis set [equation \eqref{eq:weemu}], whereas the diverging Coulomb electron-electron interaction expanded in a finite basis set is involved in the definition of $\psibasis$. 
Therefore, as these two wave functions qualitatively represent the same type of physics, a possible way to link $\efuncbasis$ and $\ecmubis $ is to try to map the projection of the diverging Coulomb interaction on a finite basis set to a non-diverging long-range effective interaction.

\subsection{Effective Coulomb electron-electron interaction in a finite basis set}
This section introduces a real-space representation of the Coulomb electron-electron operator projected in a basis set $\basis$, which is needed to derive approximations for $\efuncbasis$. 
\label{sec:real_Coulomb}
\subsubsection{Expectation values over the Coulomb electron-electron operator}
\label{sec:expectation_value}
The Coulomb electron-electron operator restricted to a basis set $\basis$ is most naturally written in orbital-space second quantization:
\begin{equation}
   \begin{aligned}
     \weeopbasis =  \frac{1}{2}\,\, \sum_{ijkl\,\,\in\,\,\basis} \,\, \vijkl \,\, \aic{k}\aic{l}\ai{j}\ai{i},
   \end{aligned}
\end{equation}
where the sums run over all (real-valued) orthonormal spin-orbitals $\{\phi_i \}$ in the basis set $\basis$, and $\vijkl$ are the two-electron integrals. 
By expanding the creation and annihilation operators in terms of real-space creation and annihilation field operators, the expectation value of $\weeopbasis$ over a wave function $\psibasis$ can be written as (see Appendix \ref{sec:appendix_expectation} for a detailed derivation): 
\begin{equation}
  \elemm{\psibasis}{\weeopbasis}{\psibasis} =  \frac{1}{2}\,\,\iint \dr{1}\,\dr{2} \,\, \fbasis,
\end{equation}
where we introduced the function
\begin{equation}
  \label{eq:fbasis}
  \begin{aligned}
  \fbasis =  \sum_{ijklmn\,\,\in\,\,\basis} & \vijkl \,\, \gammaklmn{\psibasis} \\& \phix{n}{2} \phix{m}{1}  \phix{i}{1} \phix{j}{2},
  \end{aligned}
\end{equation}
and $\gammamnpq{\psibasis}$ is the two-body density matrix of $\psibasis$
\begin{equation}
  \gammamnpq{\psibasis} = \elemm{\psibasis}{ \aic{p}\aic{q}\ai{n}\ai{m} }{\psibasis},
\end{equation}
and $\bfr{}$ collects the space and spin variables.  
\begin{equation}
 \label{eq:define_x}
  \begin{aligned}
 &\bfr{} = \left({\bf r},\sigma \right)\qquad {\bf r} \in {\rm I\!R}^3,\,\, \sigma = \pm \frac{1}{2}\\
 &\int \, \dr{} = \sum_{\sigma = \pm \frac{1}{2}}\,\int_{{\rm I\!R}^3} \, \text{d}{\bf r}.
  \end{aligned}
\end{equation}
From the properties of the restriction of an operator to the space generated by the basis set $\basis$, we have the following equality
\begin{equation}
  \elemm{\psibasis}{\weeopbasis}{\psibasis} = \elemm{\psibasis}{\weeop}{\psibasis},
\end{equation}
which translates into
\begin{equation}
  \begin{aligned}
    \label{eq:int_eq_basis}
    & \frac{1}{2}\,\,\iint  \dr{1}\,\dr{2} \,\,\fbasis \\ =\,\,&\frac{1}{2}\,\,\iint \dr{1}\,\dr{2} \,\,\frac{1}{\norm{\bfrb{1} - \bfrb{2} }} \,\,  \twodmrdiag{1}{2}{\psibasis},
  \end{aligned}
\end{equation}
where $\twodmrdiag{1}{2}{\psibasis}$ is the pair density of $\psibasis$.
Therefore, by introducing the following function 
\begin{equation}
  \label{eq:def_weebasis}
  \wbasis = \frac{\fbasis}{\twodmrdiagpsi},
\end{equation}
one can rewrite equation \eqref{eq:int_eq_basis} as
\begin{equation}
  \begin{aligned}
    \label{eq:int_eq_wee}
    & \iint \dr{1}\,\dr{2} \,\, \wbasis \,\, \twodmrdiagpsi \\ = &\iint \dr{1}\,\dr{2} \,\,\frac{1}{\norm{\bfrb{1} - \bfrb{2} }} \,\,  \twodmrdiagpsi.
  \end{aligned}
\end{equation}
One can thus identify $\wbasis$ as an effective interaction, coming from the restriction to the basis set $\basis$. This can be seen as a generalization of the exchange potential of Slater\cite{slater_exch_potential}. It is important to notice that all the quantities appearing in the integrals of equation \eqref{eq:int_eq_wee} can be considered as functions and not operators or distributions, and therefore they can be compared pointwise. Of course, the function $\wbasis$ is not defined when $\twodmrdiagpsi$ vanishes, but we leave this for a future study. 

Equation \eqref{eq:int_eq_wee} means that the two integrands have the same integral, but it does not mean that they are equals at each point $(\bfr{1},\bfr{2})$. Of course, one could argue that there exists an infinite number of functions of $u(\bfr{1},\bfr{2})$ satisfying
\begin{equation}
  \begin{aligned}
    & \iint  \dr{1}\,\dr{2} \,\, u(\bfr{1},\bfr{2}) \,\, \twodmrdiagpsi \\ = &\iint \dr{1}\,\dr{2} \,\,\frac{1}{\norm{\bfrb{1} - \bfrb{2} }} \,\,  \twodmrdiagpsi,
  \end{aligned}
\end{equation}
which implies that the effective interaction is not uniquely defined, and that the choice of equation \eqref{eq:def_weebasis} is just one among many and might not be optimal.  
For instance, the definition of the effective electron-electron interaction of equation \eqref{eq:def_weebasis} implies that it can depend on the spin of the electrons, whereas the exact Coulomb electron-electron interaction does not.  
Nevertheless, one can show (see Appendix \ref{sec:complete_basis}) that, in the limit of a complete basis set (written as ``$\basis \rightarrow \infty$''), $\wbasis$ correctly tends to the exact Coulomb interaction: 
\begin{equation}
   \label{eq:weeb_infty}
    \lim_{\basis \rightarrow \infty} \wbasis  = \frac{1}{\norm{\bfrb{1} - \bfrb{2}}}, \quad \forall \,\, (\bfr{1},\bfr{2}) \,\, \text{and }\psibasis.
\end{equation}
In particular, in this limit, $\wbasis$ does not depend on $\psibasis$ or on the spins of the electrons.

\begin{figure*}
\includegraphics[width=0.45\linewidth]{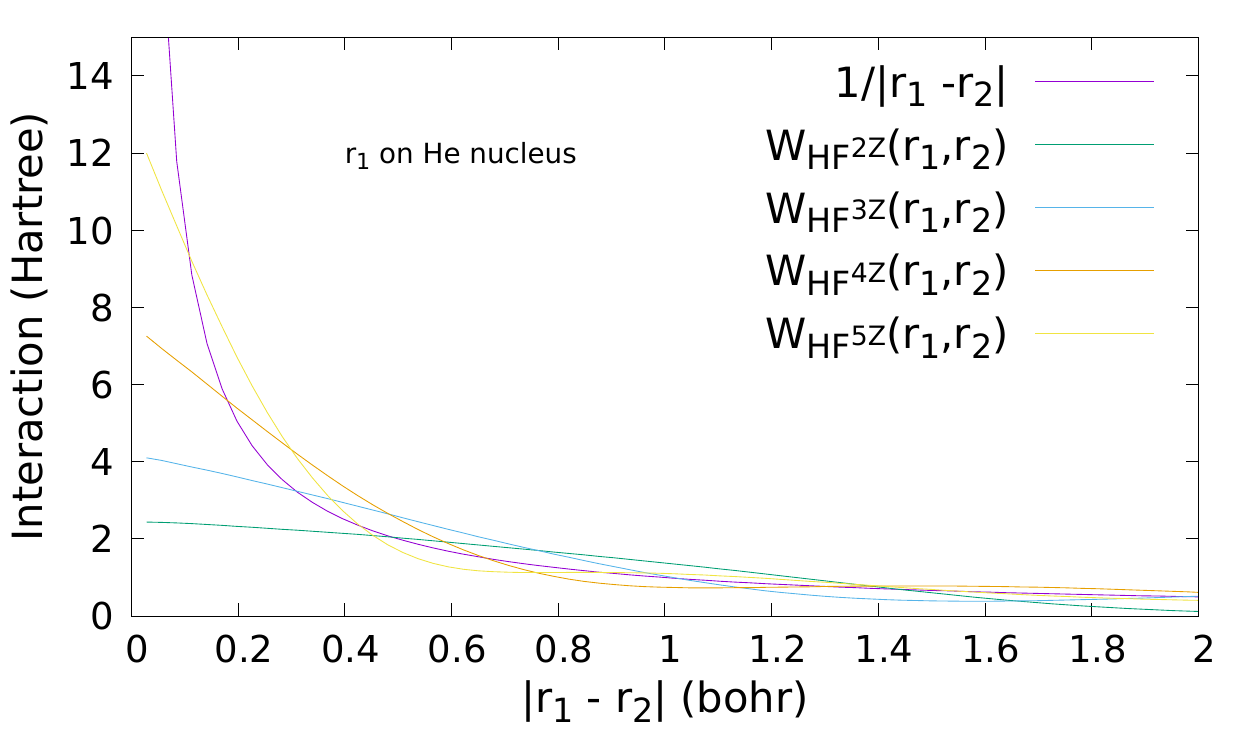}
\includegraphics[width=0.45\linewidth]{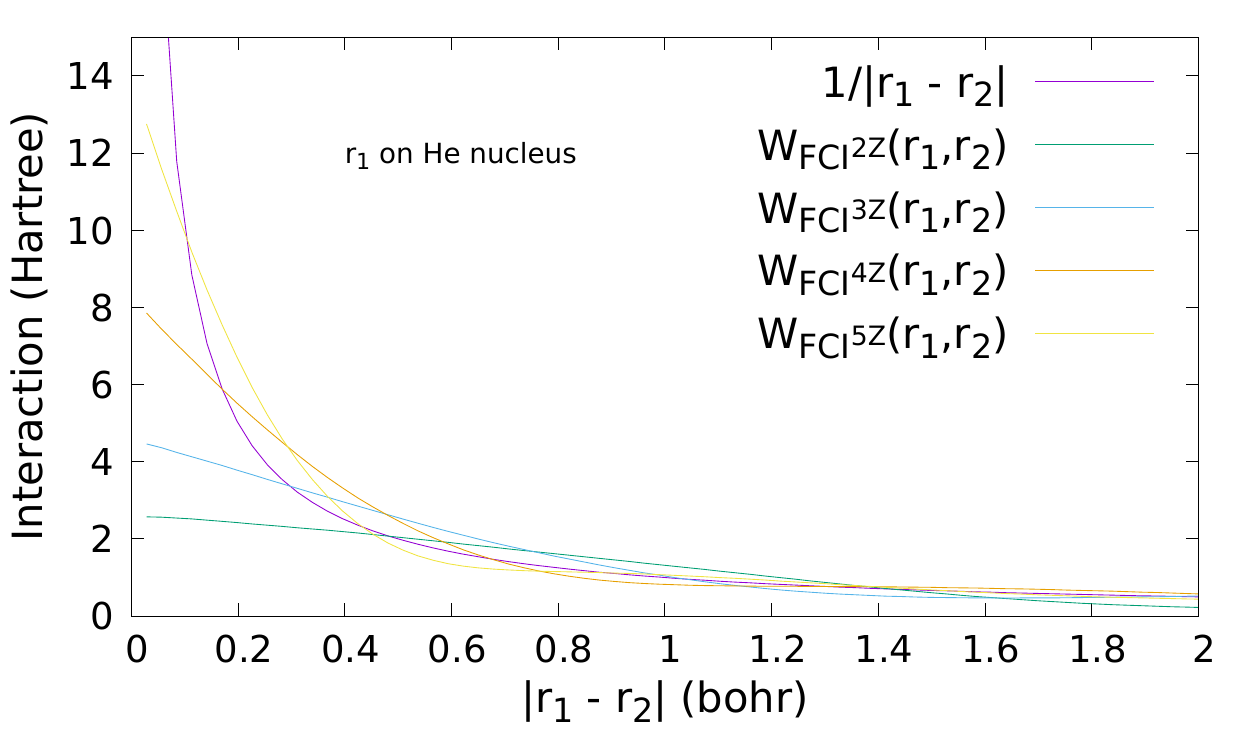}
\includegraphics[width=0.45\linewidth]{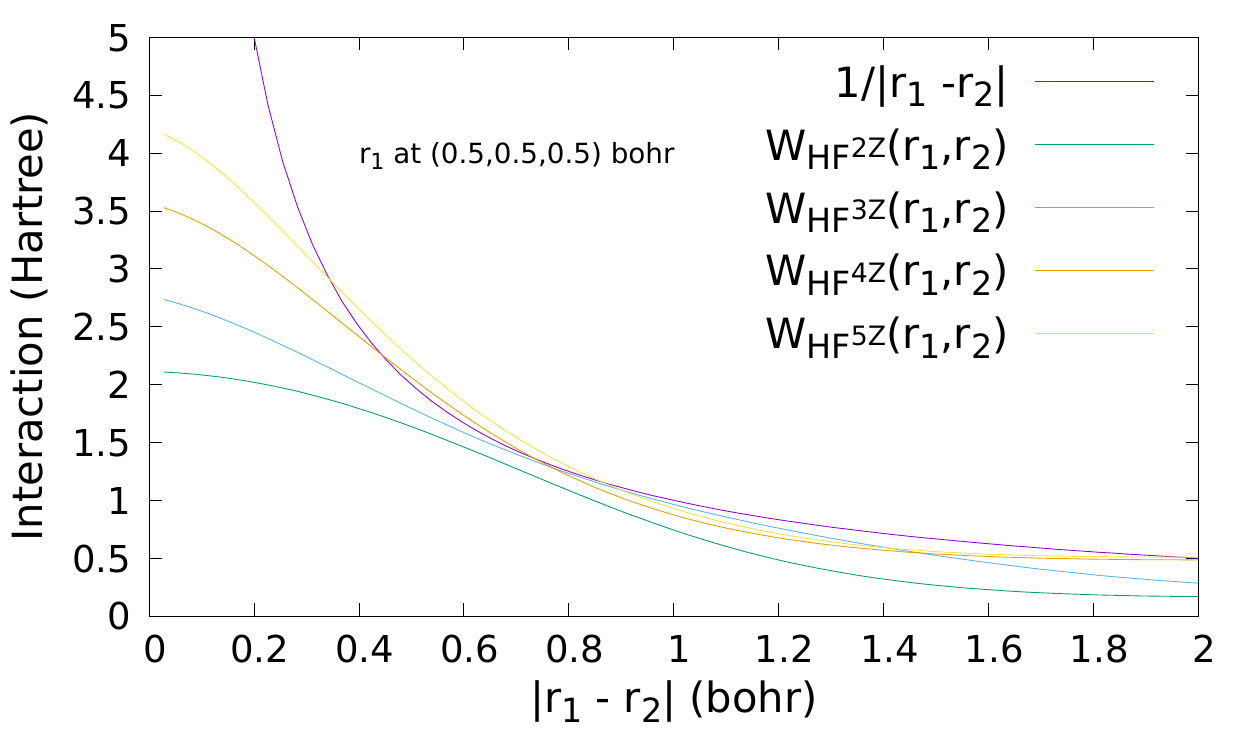}
\includegraphics[width=0.45\linewidth]{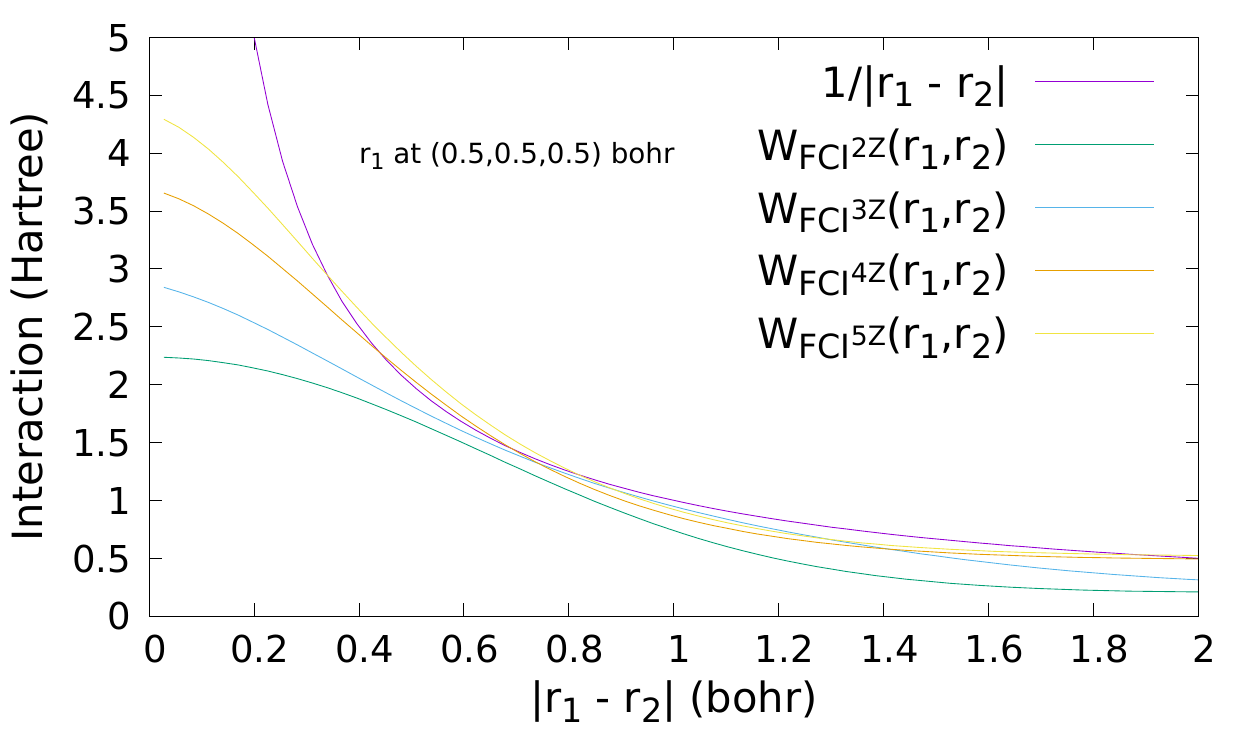}
  \caption{Effective electron-electron interaction $\wbasisr$ in the helium atom for different cc-pV$X$Z basis sets ($X=2,3,4,5$) as a function of $|{\bf r}_1 - {\bf r}_2 |$. The two upper curves are for a reference point ${\bf r}_1$ at the helium nucleus and ${\bf r}_2$ moving along the diagonal of the $xy$ plane, and the two lower curves are for a reference point ${\bf r}_1$ at $(0.5,0.5,0.5)$ bohr from the helium nucleus and ${\bf r}_2$ moving along the diagonal of the $xy$ plane with $z=0.5$ bohr. Two types of wave functions $\psibasis$ have been used: HF and FCI in the corresponding basis set. The exact Coulomb interaction ${1}/{|{\bf r}_1 - {\bf r}_2 |}$ is also reported for comparison. }
\label{fig:centered}
\end{figure*}

\subsubsection{Effective electron-electron interaction for opposite spins $\wbasisr$ and its properties}
\label{sec:trncated_basis}
The fact that $\wbasis$ tends to the exact Coulomb electron-electron interaction in the complete-basis-set limit supports the choice of this effective interaction. Nevertheless, it is also important to analyze a few properties of $\wbasis$ in the finite basis sets used in actual quantum chemistry calculations, and to understand how it differs from the true interaction.

We will consider the effective electron-electron interaction between electrons of opposite spins ($\sigma$ and $\bar{\sigma}$)
\begin{equation}
  W_{\psibasis}(\bfrb{1},\bfrb{2}) = W_{\psibasis}(\bfrb{1}\sigma,\bfrb{2}\bar{\sigma}),
\end{equation}
since the interaction between same-spin electrons is normally not the limiting factor for basis convergence.
The first thing to notice is that, because in practice $\basis$ is composed of atom-centered basis functions, the effective interaction $\wbasisr$ is not translationally invariant nor isotropic, which means that its does not depend only on the variable $|\bfrb{1}-\bfrb{2}|$
\begin{equation}
  \label{eq:wbasis_iso}
  W_{\psibasis}(\bfrb{1},\bfrb{2}) \ne W_{\psibasis}(|\bfrb{1}-\bfrb{2}|). 
\end{equation}
Thus, the quality of the representation of the Coulomb electron-electron operator (and therefore of the electron correlation effects) are not expected to be spatially uniform. Nevertheless, $\wbasisr$ is symmetric in $\bfrb{1}$ and $\bfrb{2}$:
\begin{equation}
  \wbasisr = W_{\psibasis}(\bfrb{2},\bfrb{1}).
\end{equation}

A simple but interesting quantity is the value of the effective interaction $\wbasisr$ at coalescence at a given point in space ${\bf r}_1$
\begin{equation}
 \label{eq:def_wcoal}
 \wbasiscoal{1} = W_{\psibasis}(\bfrb{1},\bfrb{1}).
\end{equation}
In a finite basis set, $\fbasis $ is finite as it is obtained from a finite sum of bounded quantities [see equation \eqref{eq:fbasis}]. Therefore, provided that the on-top pair density does not vanish, $\ontop{\psibasis}{r} = n^{(2)}_{\psibasis}(\bfrb{1}\sigma,\bfrb{1}\sigma') \ne 0$, $\wbasiscoal{1}$ is necessarily finite in a finite basis set: 
\begin{equation}
  \begin{aligned}
   \label{eq:wbasis_coalescence_1}
     \wbasiscoal{1} < \infty, \qquad \forall \,\, {\bf r}_1 \,\, \text{such that} \,\, \ontop{\psibasis}{r} \ne 0.
  \end{aligned}
\end{equation}
As mentioned above, since the effective interaction is not translationally invariant, the value $\wbasiscoal{1}$ has no reason to be independent of ${\bf r}_1$.

\subsubsection{Illustrative examples of $\wbasisr$ on the helium atom}
\label{sec:wee_practical}
In order to investigate how $\wbasisr$ behaves as a function of the basis set, the wave function $\psibasis$, and the spatial variables $({\bf r}_1,{\bf r}_2)$, we performed calculations using Dunning basis sets of increasing sizes (from aug-cc-pVDZ to aug-cc-pV5Z) using a HF or a FCI wave function for $\psibasis$ and different reference points ${\bf r}_1$. 
We report these numerical results in figure \ref{fig:centered}.

From figure \ref{fig:centered}, several trends can be observed. First, for all wave functions $\psibasis$ and reference points ${\bf r}_1$ used here, the value of $\wbasisr$ at coalescence is finite, which numerically illustrates equation \eqref{eq:wbasis_coalescence_1}. Second, the value at coalescence increases with the cardinal of the basis set, suggesting that the description of the short-range part of the interaction is improved by enlarging the basis set. Third, the global shape of the $\wbasisr$ is qualitatively modified by changing the reference point ${\bf r}_1$, which illustrates the lack of transitional invariance of $\wbasisr$. In particular, the values of $\wbasishf$ and $\wbasisfci$ at coalescence are much larger when the reference point ${\bf r}_1$ is on the He nucleus, which is a signature that the atom-centered basis set does not uniformly describe the Coulomb interaction at all points in space.
Fourth, the difference between the $\wbasishf$ and $\wbasisfci$ is almost unnoticeable for all basis sets and for the two reference points ${\bf r}_1$ used here. 

\begin{figure*}
\includegraphics[width=0.45\linewidth]{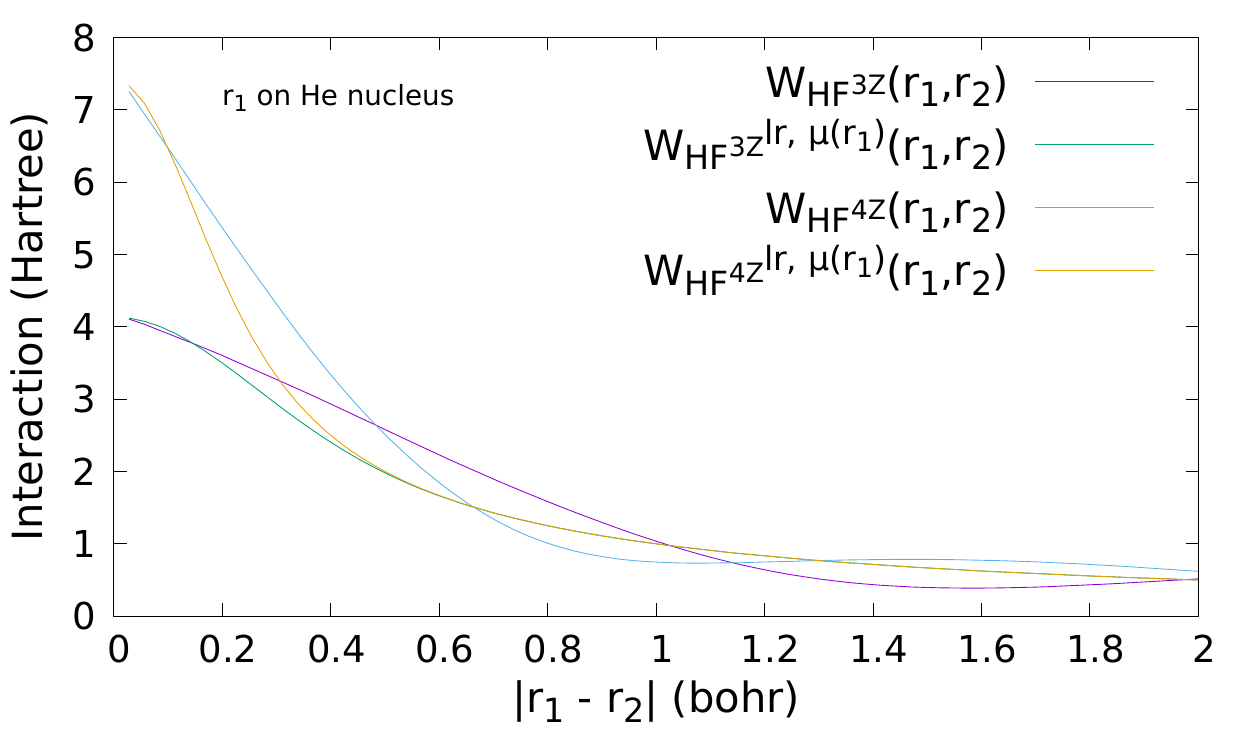}
\includegraphics[width=0.45\linewidth]{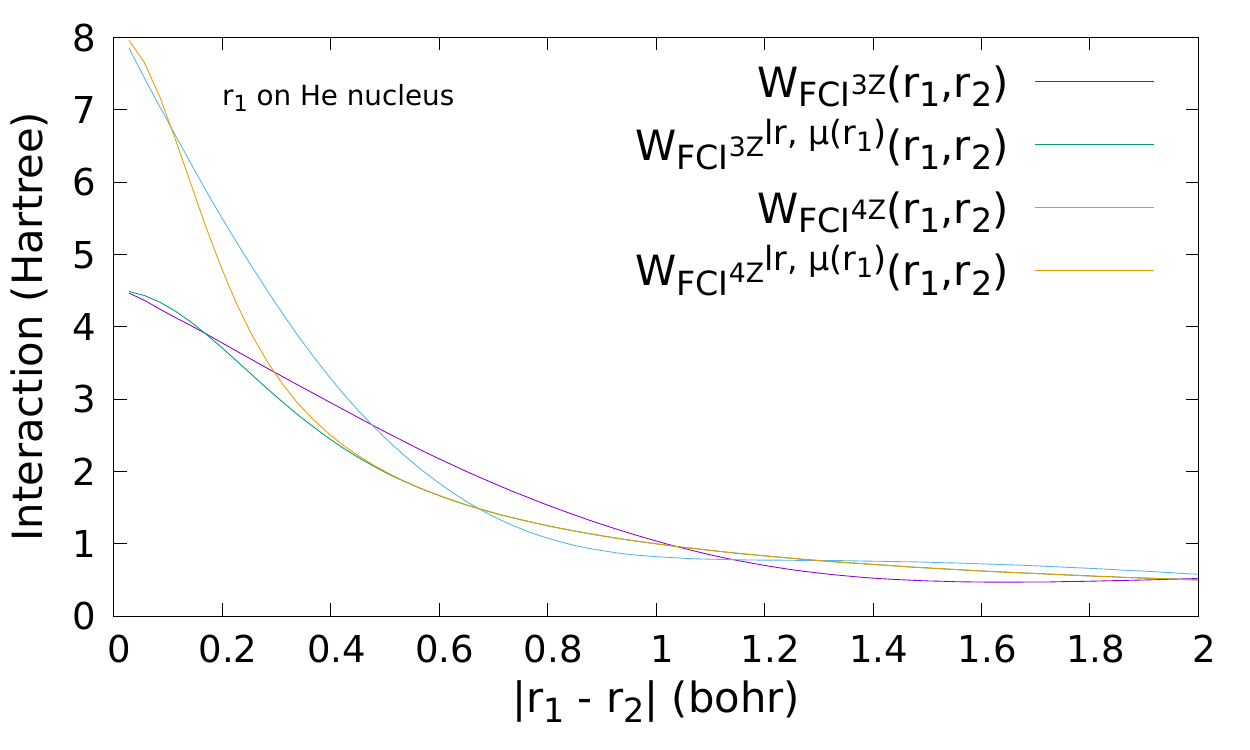}
\includegraphics[width=0.45\linewidth]{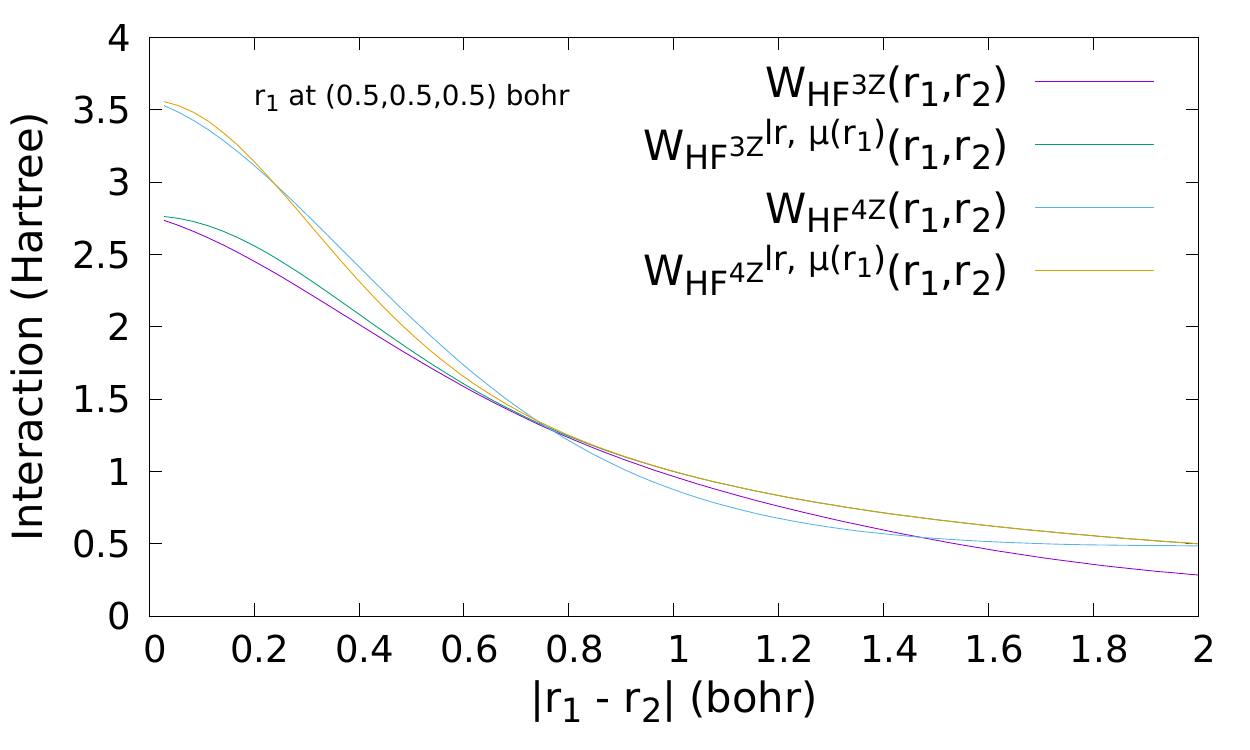}
\includegraphics[width=0.45\linewidth]{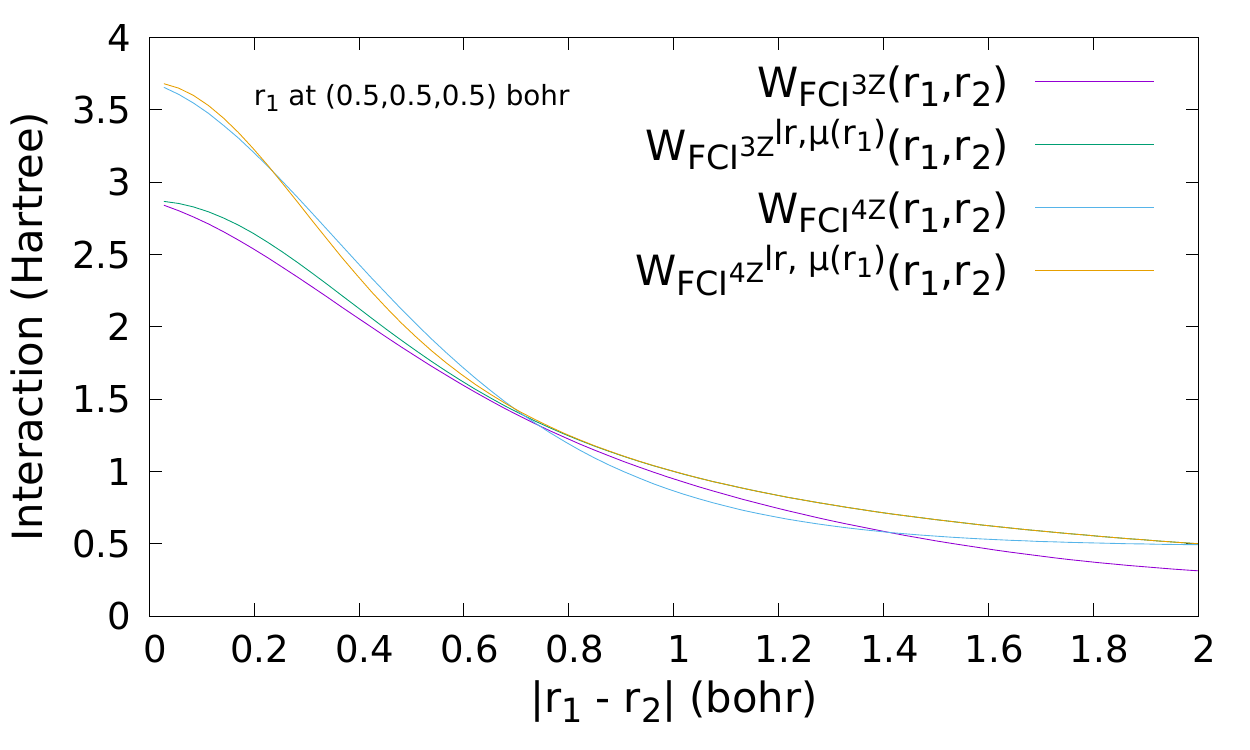}
  \caption{Effective electron-electron interaction $\wbasisr$ and long-range electron-electron interaction $\wbasismu$ for different cc-pV$X$Z basis sets ($X=3,4$) as a function of $|{\bf r}_1 - {\bf r}_2 |$. The two upper curves are for a reference point ${\bf r}_1$ at the helium nucleus and ${\bf r}_2$ moving along the diagonal of the $xy$ plane, and the two lower curves are for a reference point ${\bf r}_1$ at $(0.5,0.5,0.5)$ bohr from the helium nucleus and ${\bf r}_2$ moving along the diagonal of the $xy$ plane with $z=0.5$ bohr. Two types of wave functions $\psibasis$ have been used: HF and FCI in the corresponding basis set. }
\label{fig:erf_centered}
\end{figure*}

\subsubsection{Link with range-separated DFT: Introduction of a local range-separated parameter $\mu({\bf r})$}
\label{sec:mu_of_r}
From the numerical illustration of the properties of $\wbasisr$ given in section \ref{sec:wee_practical}, it appears that the development of approximations for the density functional $\efuncbasis$ seems rather complicated since the effective interaction $\wbasisr$ is system- and basis-dependent, non translationally invariant, and non isotropic. Nevertheless, as it was numerically illustrated, the effective interaction $\wbasisr$ typically describes a long-range interaction which is finite at coalescence. 
Therefore, a possible way to approximate $\wbasisr$ is to locally fit $\wbasisr$ by the long-range interaction $w^{\text{lr},\mu}(|{\bf r}_1 -{\bf r}_{2}|)$ of equation \eqref{eq:erf} used in RS-DFT.
To do so, we propose here to determine a local value of the range-separation parameter $\mu$ such that the value of the long-range interaction at coalescence is identical to the value of the effective interaction $\wbasiscoal{1}$ at coalescence at point ${\bf r}_1$. More specifically, the range-separation parameter $\mu({\bf r}_1;\psibasis)$ is thus determined for each ${\bf r}_1$ and $\psibasis$ by the condition:
\begin{equation}
  \wbasiscoal{1} = w^{\text{lr},\mu({\bf r}_1;\psibasis)}(0),
\end{equation}
with $\wbasiscoal{1}$ given by equations \eqref{eq:def_wcoal} which, since $w^{\text{lr},\mu}(0) = 2\mu/\sqrt{\pi}$, simply gives
\begin{equation}
 \label{eq:mu_of_r}
 \mu({\bf r}_1;\psibasis) = \frac{\sqrt{\pi}}{2} \, \wbasiscoal{1} \, .
\end{equation}
Therefore, defining the function $\wbasismu$ as
\begin{equation}
 \begin{aligned}
 \label{eq:wee_mu_of_r}
  \wbasismu = \frac{\text{erf}\left(\mu({\bfrb{1}}; \psibasis)|\bfrb{1} - \bfrb{2}|\right)}{|\bfrb{1} - \bfrb{2}|} \, , 
 \end{aligned}
\end{equation}
we make the following approximation:
\begin{equation}
 \label{eq:wbasis_approx}
 \wbasisr \approx \wbasismu, \qquad \forall \,\,\, (\bfrb{1},\bfrb{2}) \, .
\end{equation}
One can notice that the definition of $\mu({\bf r}_1;\psibasis)$ in equation \eqref{eq:mu_of_r} depends on the choice of $\psibasis$, and therefore the approximation of equation \eqref{eq:wbasis_approx} depends also on $\psibasis$. Nevertheless, in the limit of a complete basis set the dependence on $\psibasis$ vanishes. 

\begin{figure*}[t]
\includegraphics[width=0.45\linewidth]{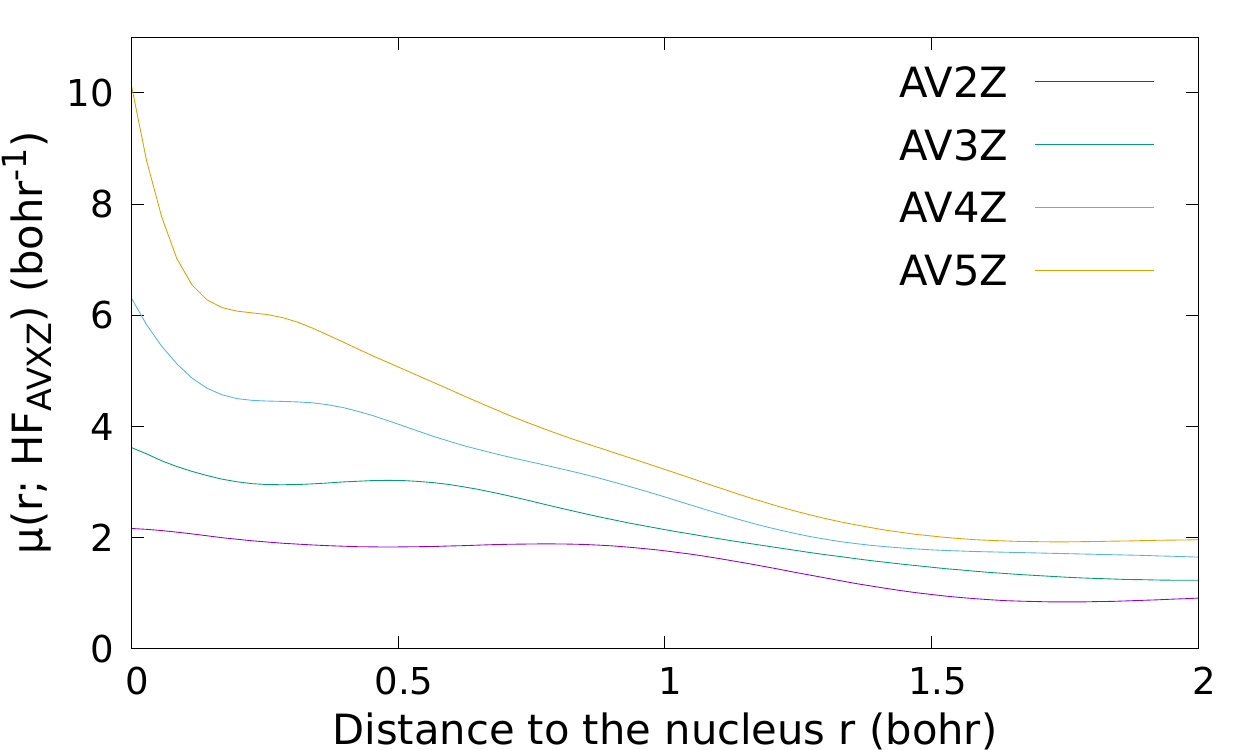}
\includegraphics[width=0.45\linewidth]{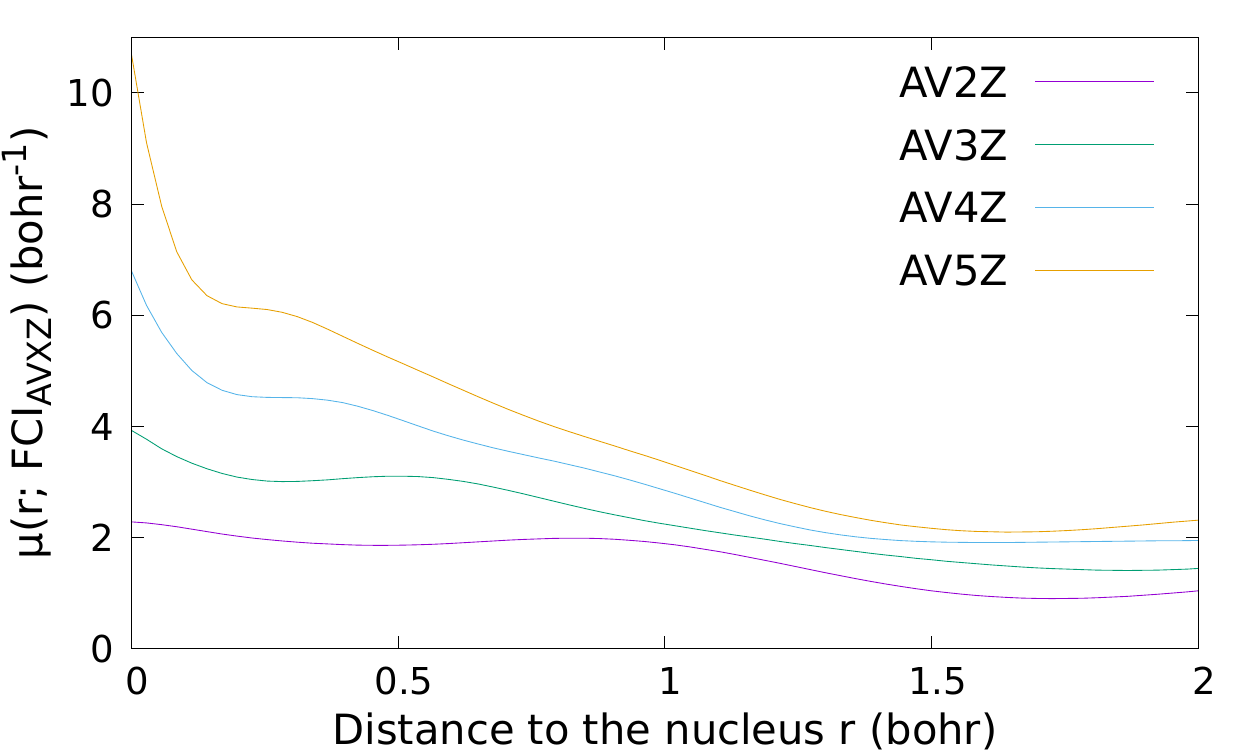}
  \caption{Local range-separated parameter $\mu({\bf{r}};\psibasis)$ for the helium atom for different aug-cc-pV$X$Z basis sets ($X=2,3,4,5$) as a function of the position ${\bf r}$ along the diagonal of the $xy$ plane. The curve on the left is when using the HF wave function for $\psibasis$, and the curve on the right is when using the FCI wave function for $\psibasis$.}
\label{fig:mu_of_r}
\end{figure*}

In order to illustrate how $\wbasismu$ compares to $\wbasisr$, we report in figure \ref{fig:erf_centered} these two functions for several basis sets, for different reference points ${\bf r}_1$, and for two different wave functions $\psibasis$. From these plots it appears that the approximation of equation \eqref{eq:wbasis_approx} is reasonably accurate when the reference point ${\bf r}_1$ is on the helium nucleus and becomes even more accurate when the reference point ${\bf r}_1$ is farther away from the helium nucleus.  

In figure \ref{fig:mu_of_r}, we report the local range-separation parameter $\mu({\bf{r}};\psibasis)$, as determined by equation \eqref{eq:mu_of_r}, for different basis sets and when $\psibasis$ is the HF or FCI wave function. It clearly appears that the magnitude of $\mu({\bf{r}};\psibasis)$ increases when the size of the basis set increases, which translates the fact that the electron-electron interaction is better described by enlarging the basis set. Also, for all basis sets, the maximal value of $\mu({\bf{r}};\psibasis)$ is reached when ${\bf r}$ is at the nucleus, which demonstrates the non-homogeneity of the description of the electron-electron interaction with atom-centered basis functions. Finally, one can notice that the values of $\mu({\bf{r}};\psibasis)$ are very similar when using the HF or FCI wave function for $\psibasis$, but nevertheless slightly larger for the FCI wave function which reflects the fact that the corresponding effective interaction is slightly stronger.

\subsection{Practical approximations for the complementary functional $\efuncbasis$: a short-range LDA-like functional with a local $\mu({\bf r})$}
\label{sec:practical_approx}
A proper way to define an LDA-like approximation for the complementary density functional $\efuncbasis$ would be to perform a uniform-electron gas calculation with the function $\wbasisr$ as the electron-electron interaction. However, such a task would be rather difficult and ambiguous as $\wbasisr$ is not translationally invariant nor isotropic, which thus questions how a uniform density could be obtained from such an interaction. Instead, by making the approximation of equation \eqref{eq:wbasis_approx}, one can define for each point $\bfrb{1}$ an effective interaction which only depends on $|\bfrb{1}-\bfrb{2}|$. For a given point in space ${\bf r}_1$ one can therefore use the multi-determinant short-range correlation density functional of equation \eqref{eq:ec_md_mu} with the range-separation parameter value $\mu({\bf r}_1;\psibasis)$ corresponding to a local effective interaction at ${\bf r}_1$ [see equation \eqref{eq:mu_of_r}]. 
Therefore, we define an LDA-like functional for $\efuncbasis$ as
\begin{equation}
 \label{eq:def_lda_tot}
  \efuncbasislda = \int \, \text{d}{\bf r} \,\, \denrbasis \,\, \emulda\,,
\end{equation}
where $\bar{\varepsilon}^{\text{sr},\text{unif}}_{\text{c,md}}(n,\mu)$ is the multi-determinant short-range correlation energy per particle of the uniform electron gas for which a parametrization can be found in Ref.~\onlinecite{PazMorGorBac-PRB-06}. In practice, for open-shell systems, we use the spin-polarized version of this functional (i.e., depending on the spin densities) but for simplicity we will continue to use only the notation of the spin-unpolarized case.
One can interpret equation \eqref{eq:def_lda_tot} as follows: the total correction to the energy in a given basis set is approximated by the sum of local LDA corrections obtained, at each point, from an uniform electron gas with a specific electron-electron interaction which approximatively coincides with the local effective interaction obtained in the basis set. 
Within the LDA approximation, the final working equation for our basis-correction scheme is thus
\begin{equation}
 \label{eq:def_work}
  \emuldafinal = E_{\text{FCI}}^\basis + \efuncbasisldafci.
\end{equation}
We will refer to this approach as FCI+LDA$_{\psibasis}$ where $\psibasis$ indicates the wave function used to define the effective interaction within the basis set $\basis$ employed in the calculation.

\subsection{Basis-set-corrected CIPSI: the CIPSI+LDA$_{\psibasis}$ approach}
\label{sec:cipsi_lda}
Equation \eqref{eq:def_work} requires the calculation of the FCI energy and density whose computational cost can be very rapidly prohibitive. In order to remove this bottleneck, we propose here a similar approximation to correct the so-called CIPSI energy which can be used to approximate the FCI energy in systems where the latter is out of reach.

\subsubsection{The CIPSI algorithm in a nutshell}
The CIPSI algorithm approximates the FCI wave function through an iterative selected CI procedure, and the FCI energy through a second-order multi-reference perturbation theory. 
The CIPSI algorithm belongs to the general class of methods build upon selected CI\cite{bender,malrieu,buenker1,buenker-book,three_class_CIPSI,harrison,hbci}
which have been successfully used to converge to FCI correlation energies, one-body properties, and nodal surfaces.\cite{three_class_CIPSI,Rubio198698,cimiraglia_cipsi,cele_cipsi_zeroth_order,Angeli2000472,canadian,atoms_3d,f2_dmc,atoms_dmc_julien}
The CIPSI algorithm used in this work uses iteratively enlarged selected CI spaces and 
Epstein--Nesbet\cite{epstein,nesbet} multi-reference perturbation theory. The CIPSI energy is
\begin{align}
  E_\mathrm{CIPSI} &= E_\text{v} + E^{(2)},
\end{align}
where $E_\text{v}$ is the variational energy
\begin{align}
  E_\text{v} &= \min_{\{ c_{\rm I}\}} \frac{\elemm{\Psi^{(0)}}{\hat{H}}{\Psi^{(0)}}  }{\ovrlp{\Psi^{(0)}}{\Psi^{(0)}}},
\end{align}
where the reference wave function $\ket{\Psi^{(0)}} = \sum_{{\rm I}\,\in\,\mathcal{R}} \,\,c_{\rm I} \,\,\ket{\rm I}$ is expanded in Slater determinants I within the CI reference space $\mathcal{R}$, and $E^{(2)}$ is the second-order energy correction
\begin{align}
  E^{(2)} &= \sum_{\kappa} \frac{|\elemm{\Psi^{(0)}}{\hat{H}}{\kappa}|^2}{E_\text{v} - \elemm{\kappa}{H}{\kappa}} = \sum_{\kappa} \,\, e_{\kappa}^{(2)},
\end{align}
where $\kappa$ denotes a determinant outside $\mathcal{R}$.  
To reduce the cost of the evaluation of the second-order energy correction, the semi-stochastic multi-reference approach
of Garniron \textit{et al.} \cite{stochastic_pt_yan} was used, adopting the technical specifications recommended in that work. 
The CIPSI energy is systematically refined by doubling the size of the CI reference space at each iteration, selecting
the determinants $\kappa$ with the largest $\vert e_{\kappa}^{(2)} \vert$. The calculations are stopped when a target value of $E^{(2)}$ is reached. 

\subsubsection{Working equations for the CIPSI+LDA$_{\psibasis}$ approach}
\label{sec:cipsi_dft}
The CIPSI algorithm being an approximation to FCI, one can straightforwardly apply the DFT correction developed in this work to correct the CIPSI energy error due to the basis set. 
For a given basis set $\basis$ and a given reference wave function $\Psi^{(0)}$, one can estimate the FCI energy and density by the following approximations: 
\begin{equation}
 \label{eq:cipsifci}
 E_{\text{FCI}}^\basis \approx  E_{\text{CIPSI}}^\basis \, , 
\end{equation}
\begin{align}
 \label{eq:cipsidensity}
 \denfci({\bf r}) \approx \dencipsi  \, , 
\end{align}
with 
\begin{align}
  \dencipsi = \elemm{\Psi^{(0)}}{\hat{n}({\bf r})}{\Psi^{(0)}}  \, . 
\end{align}
Assuming these approximations, for a given choice of $\psibasis$ to define the effective interaction and within the LDA approximation of equation \eqref{eq:def_lda_tot}, one can define the corrected CIPSI energy as 
\begin{equation}
 \label{eq:E_B_fci}
   E_{\text{CIPSI+LDA}_{\psibasis}}^{\basis,\psibasis}  =  E_{\text{CIPSI}}^\basis + \efuncbasisldacipsi  \, .
\end{equation}
Note that the reference wave function $\Psi^{(0)}$ can be used for the definition of the effective interaction through its two-body density matrix [see equation \eqref{eq:fbasis}], but we leave that for further investigation and for the rest of the calculations we use the HF wave function for $\psibasis$ in the definition of the effective interaction, and we denote the method by CIPSI+LDA$_{\text{HF}}$.

\begin{figure}[t]
\includegraphics[width=\linewidth]{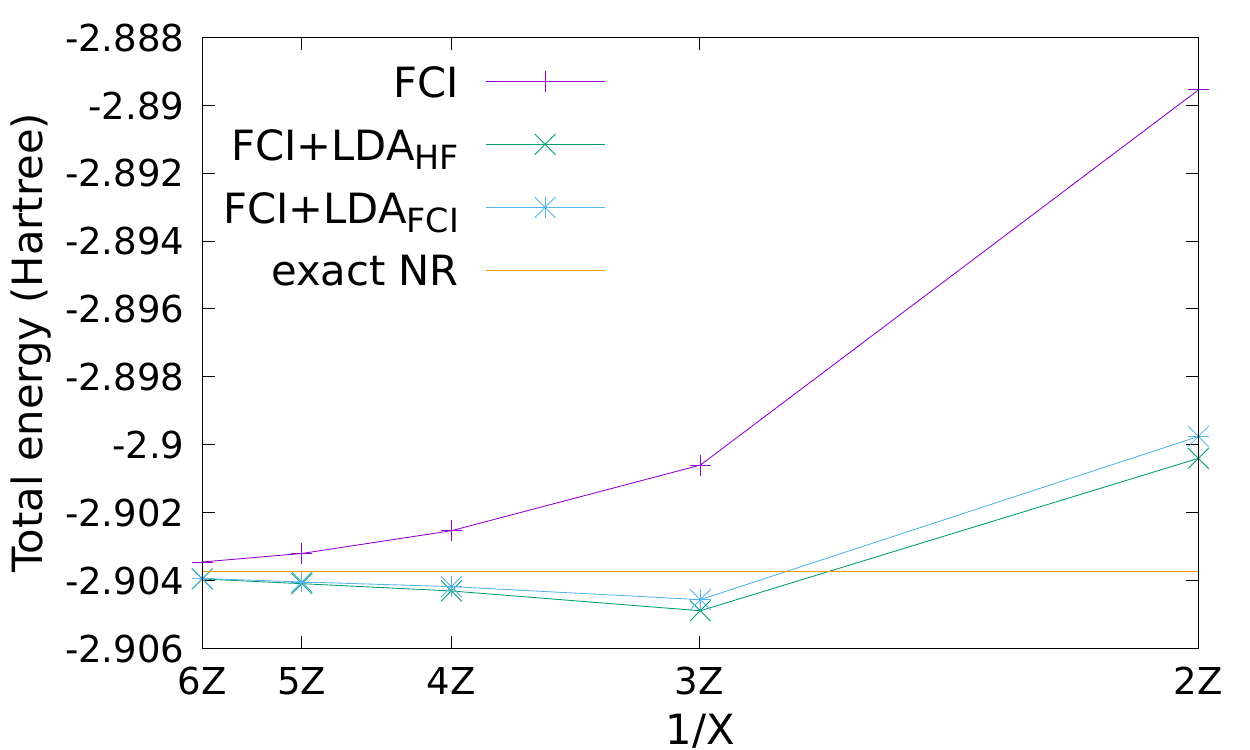}
  \caption{Convergence of the total energy of the helium atom for FCI and FCI+LDA$_{\psibasis}$, where $\psibasis$ is either the HF or FCI wave function, as a function of the inverse of the cardinal number $X$ of the AV$X$Z basis sets ($X=2,3,4,5,6$). The exact non-relativistic (NR) energy is also reported.}
\label{fig:conv_He}
\end{figure}

\begin{table*}
\caption{Total energies (in Hartree) of the helium atom and errors (in mH) with respect to the exact non-relativistic energy for FCI, FCI+LDA$_{\text{HF}}$, and FCI+LDA$_{\text{FCI}}$ with the AV$X$Z basis sets ($X=2, 3, 4, 5, 6$). }
\begin{ruledtabular}
\begin{tabular}{ldddddd}
                        & \multicolumn{2}{c}{{FCI}}     & \multicolumn{2}{c}{{FCI+LDA$_{\text{HF}}$}}  &    \multicolumn{2}{c}{{FCI+LDA$_{\text{FCI}}$}}       \\
\cline{2-3} \cline{4-5} \cline{6-7}
                        & \mc{Total energy}   & \mc{Error}       & \mc{Total energy}   & \mc{Error}                      & \mc{Total energy}   & \mc{Error}                                \\
\hline
AV2Z             & -2.88955      &   14.17     &  -2.90040     &  3.3187                    & -2.89976      &   3.962                      \\
AV3Z             & -2.90060      &   03.12     &  -2.90489     & -1.1698                    & -2.90456      &  -0.840                      \\ 
AV4Z             & -2.90253      &   01.18     &  -2.90430     & -0.5849                    & -2.90418      &  -0.460                      \\
AV5Z             & -2.90320      &   00.52     &  -2.90409     & -0.3710                    & -2.90404      &  -0.321                      \\ 
AV6Z             & -2.90346      &   00.26     &  -2.90396     & -0.2367                    & -2.90394      &  -0.217                      \\[0.2cm]
 & \multicolumn{6}{c}{Exact non-relativistic total energy} \\
 & \multicolumn{6}{c}{ -2.90372} \\
\end{tabular}
\end{ruledtabular}
\label{conv_He_table}
\end{table*}

\section{Numerical tests: Total energy of He and ionization potentials for the B-Ne atomic series}
\label{sec:numerical}
For the present study, we use the LDA approximation of equation \eqref{eq:def_lda_tot} and investigate the convergence of the total energies and energy differences as a function of the basis set. All calculations were performed with Quantum Package\cite{qp} using the Dunning aug-cc-pV$X$Z basis sets which are referred here as AV$X$Z. 

\subsection{FCI+DFT: Total energy of the helium atom}
\label{sec:fci_numerical}
We report in figure \ref{fig:conv_He} and table \ref{conv_He_table} the convergence of the total energies computed for the helium atom in the Dunning basis sets AV$X$Z ($X=2, 3, 4, 5, 6$) using FCI and  FCI+LDA$_{\psibasis}$ where $\psibasis$ is either the HF or FCI wave function. The first striking observation from these data is that the FCI+LDA$_{\psibasis}$ energies rapidly converge to the exact energy as one increases the size of the basis set and that FCI+LDA$_{\psibasis}$ is systematically closer to the exact energy than the FCI energy. Also, one can observe that $\efuncbasislda$ overestimates the correlation energy (in absolute value) for the AV3Z basis and the larger ones, which is consistent with the fact that LDA is known to give too negative correlation energies in regular Kohn-Sham DFT or in RS-DFT. Interestingly, $\efuncbasislda$ is almost independent of the choice of the wave function $\psibasis$ used for the definition of the effective interaction within $\basis$, as the FCI+LDA$_{\text{HF}}$ and FCI+LDA$_{\text{FCI}}$ energies are overall very close and get closer as one increases the size of the basis set. This last point is the numerical illustration that, in the limit of a complete basis set, the effective interaction is independent of the wave function $\psibasis$ [see equation \eqref{eq:weeb_infty}]. Nevertheless, one observes that the correction obtained using the FCI wave function for $\psibasis$ is systematically smaller in absolute value than the one obtained with the HF wave function for $\psibasis$. This result can be qualitatively understood by noticing that the introduction of the HF two-body density matrix in equation \eqref{eq:fbasis} reduces the number of two-electron integrals involved in the definition of $\wbasis$ [see equation \eqref{eq:def_weebasis}]. This reduction implies that the effective interaction $\wbasispsi{\text{HF}}$ misses a part of the interaction within the basis set, namely the repulsion between electrons in virtual orbitals. However, the fact that $\efuncbasisldapsi{HF}$ and $\efuncbasisldapsi{FCI}$ are close suggests that $\efuncbasisldapsi{HF}$ misses only a small part of the interaction. This statement can be intuitively understood by noticing that some two-electron integrals involved in the definition of $\wbasispsi{\text{HF}}$ are of the type $V_{ij}^{ab}$ (where $i,j$ and $a,b$ run over the occupied and virtual orbitals, respectively) which are the ones giving rise to the dominant part of the MP2 correlation energy in a given basis set.

\subsection{CIPSI+LDA: Total energies and energy differences for atomic systems}
\label{sec:cipsi_numerical}
\subsubsection{Convergence of the CIPSI+LDA$_{\text{HF}}$ total energy with the number of determinants}
\begin{figure*}[t]
\includegraphics[width=0.45\linewidth]{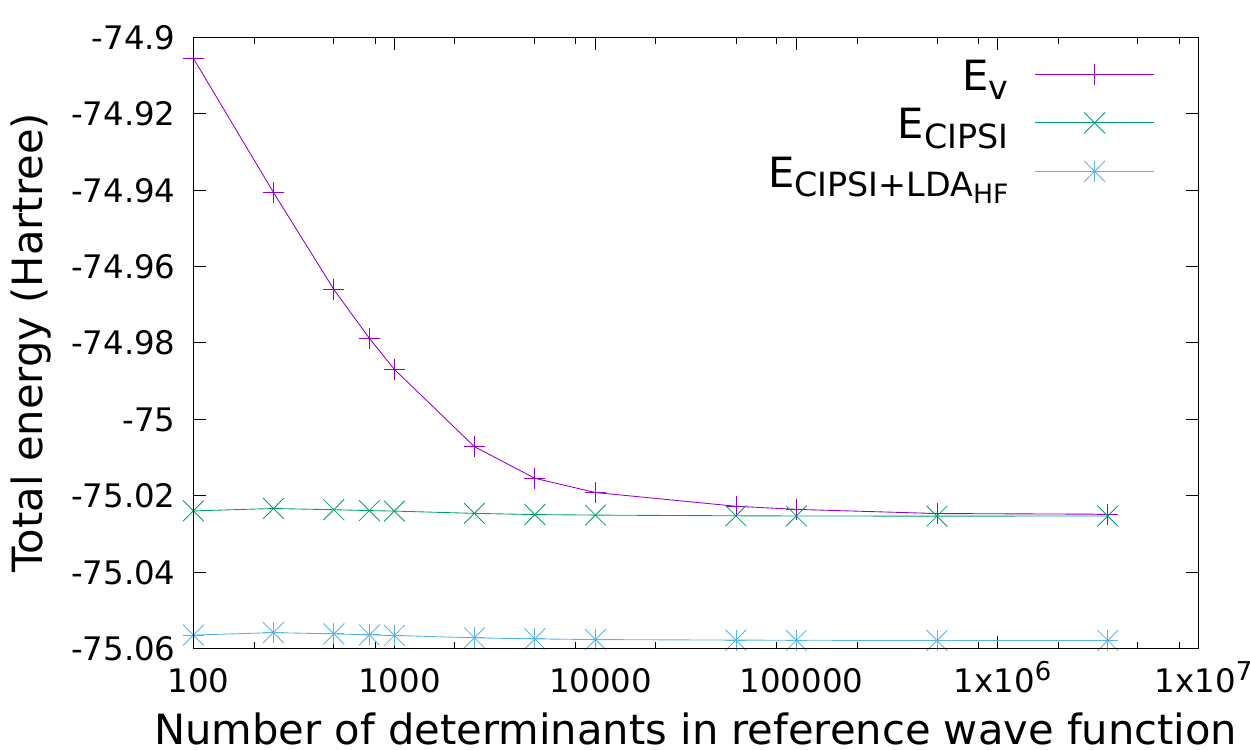}
\includegraphics[width=0.45\linewidth]{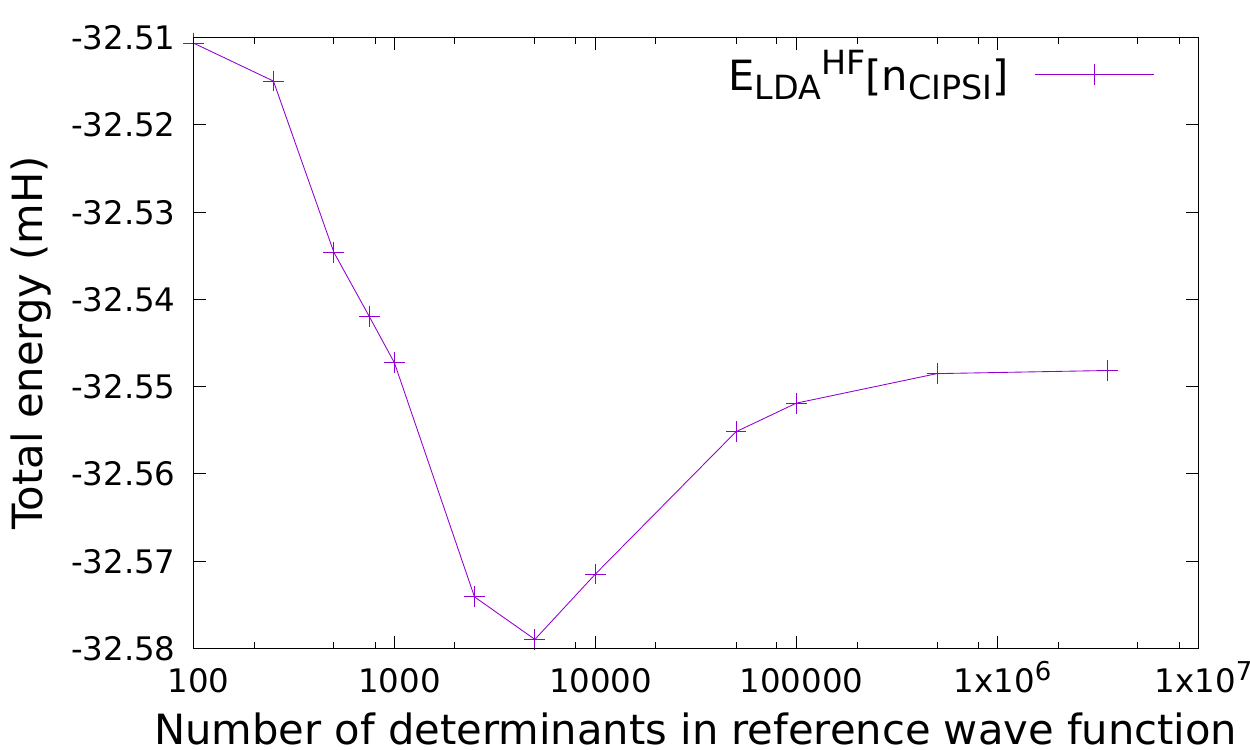}
  \caption{Convergence of the variational total energy $E_\text{v}$, the CIPSI total energy, and the CIPSI+LDA$_{\text{HF}}$ total energy (left plot), and of the LDA correction $\efuncbasisldacipsihf$ (right plot) of the oxygen atom as a function of the number of Slater determinants in the reference wave function using the AV4Z basis set. }
\label{fig:conv_cipsi}
\end{figure*}

We report in figure \ref{fig:conv_cipsi}, in the case of the oxygen ground state using the AV4Z basis set, the convergence of the variational energy $E_\text{v}$, the CIPSI energy, the CIPSI+LDA$_{\text{HF}}$ energy, and the LDA correction $\efuncbasisldacipsihf$ as a function of the number of Slater determinants in the reference wave function. The behavior of $E_\text{v}$ and $E_{\text{CIPSI}}$ reported in figure \ref{fig:conv_cipsi} are typical of a CIPSI calculation: a rapid convergence of the variational energy and an even faster convergence of the CIPSI energy. In this case, $E_{\text{CIPSI}}$ with a reference wave function including $2\times10^3$ and $5\times10^5$ determinants provides an estimation of the FCI energy with an error smaller than 1 mH and 0.1 mH, respectively, whereas the size of the FCI space of this system for this basis set is approximatively of $10^{11}$ determinants. 
Regarding $\efuncbasisldacipsihf$, it varies by about 0.08 mH between 100 and $4\times 10^{6}$ determinants. The very small variation of  $\efuncbasisldacipsihf$ can be qualitatively understood by noticing that, within the LDA approximation of equation \eqref{eq:def_lda_tot} and choosing a HF wave function for $\psibasis$ to define the effective interaction, $\efuncbasisldacipsihf$ only depends on the one-body density which is known to converge rapidly with the level of correlation treatment, especially for atomic systems. To conclude this part of the study, it can be stated that the convergence of the CIPSI+LDA$_{\text{HF}}$ energy is only limited by the convergence of the CIPSI algorithm itself as $\efuncbasisldacipsihf$ converges very rapidly with the quality of the wave function. 

 \begin{table*}
   \caption{Total energies (in Hartree) of the neutral atoms and first cations for the B-Ne series with the AV$X$Z basis sets ($X=2, 3, 4, 5$) using CIPSI and CIPSI+LDA$_\text{HF}$. The \textit{i}-FCIQMC values from Ref. \onlinecite{ip_ali} are also reported for comparison with CIPSI.}
\begin{ruledtabular}
 \begin{tabular}{llddddd}
          &Method                    &      \multicolumn{1}{c}{   AV2Z}   &         \multicolumn{1}{c}{AV3Z}     &  \multicolumn{1}{c}{AV4Z}            &     \multicolumn{1}{c}{AV5Z}        & \multicolumn{1}{c}{Exact NR$^a$}\\
\cline{2-3} \cline{4-5} \cline{6-7}
 \hline
   B      & \textit{i}-FCIQMC$^b$   &    -24.59242(1) &    -24.60665(2) &     -24.62407(11)&    -24.63023(2)  &     \\
          & CIPSI                    &    -24.592418   &    -24.606654   &     -24.624109   &    -24.630233    & -24.65390\\
          & CIPSI+LDA$_{\text{HF}}$  &    -24.641525   &    -24.641706   &     -24.648135   &    -24.650243    &      \\
 \hline
 B$^+$    & \textit{i}-FCIQMC$^b$   &    -24.29450(1) &    -24.30366(2) &     -24.32005(2) &    -24.32553(9)  &      \\
          & CIPSI                    &    -24.294496   &    -24.303660   &     -24.320044   &    -24.325531    & -24.34889\\
          & CIPSI+LDA$_{\text{HF}}$  &    -24.338930   &    -24.336580   &     -24.343043   &    -24.345024    &      \\
 \hline
 C        & \textit{i}-FCIQMC$^b$   &    -37.76656(1) &    -37.79163(2) &     -37.81301(2) &    -37.82001(4)  &         \\
          & CIPSI                    &    -37.766573   &    -37.791623   &     -37.813025   &    -37.820016    & -37.8450\\
          & CIPSI+LDA$_{\text{HF}}$  &    -37.824730   &    -37.830667   &     -37.838253   &    -37.840544    &       \\
 \hline
 C$^+$    & \textit{i}-FCIQMC$^b$   &    -37.35960(1) &    -37.37967(2) &     -37.39991(1) &    -37.40605(1)  &       \\
          & CIPSI                    &    -37.359602   &    -37.379703   &     -37.399932   &    -37.406342    & -37.43095\\
          & CIPSI+LDA$_{\text{HF}}$  &    -37.413086   &    -37.416631   &     -37.424109   &    -37.426321    &       \\
 \hline
 N        & \textit{i}-FCIQMC$^b$   &    -54.48881(2)&      -54.52797(1)&      -54.55423(3)&     -54.56303(2)&\\
          & CIPSI                    &  -54.488814  &  -54.527941  &  -54.554235  &  -54.563027& -54.5893\\
          & CIPSI+LDA$_{\text{HF}}$  &   -54.556940  &   -54.571576  &   -54.581128  &   -54.584048&\\
 \hline
 N$^+$    & \textit{i}-FCIQMC $^b$   &     -53.96106(10)&      -53.99535(1)&      -54.01838(1)&      -54.02865(2)&\\
          & CIPSI                    &  -53.961062  &  -53.995355  &  -54.020414  &  -54.028633& -54.0546\\
          & CIPSI+LDA$_{\text{HF}}$  &   -54.024314  &   -54.036820  &   -54.046204  &   -54.049068&\\
 \hline
 O        & \textit{i}-FCIQMC$^b$       &    -74.92772(2)&    -74.99077(4)&    -75.02534(4)&     -75.03749(6)&\\
          & CIPSI                        &  -74.927696  &  -74.990750  &  -75.025340  &  -75.037527& -75.0674\\
          & CIPSI+LDA$_{\text{HF}}$      &   -75.014946  &   -75.044685  &   -75.057889  &   -75.061639&\\
 \hline
 O$^+$    & \textit{i}-FCIQMC$^b$       &      -74.444194(6)&      -74.49701(1) &      -74.52799(4)&      -74.53869(6)&      \\
          & CIPSI                        &  -74.444191  &  -74.497018  &  -74.527968  &  -74.538630& -74.5669\\
          & CIPSI+LDA$_{\text{HF}}$      &   -74.517650  &   -74.543804  &   -74.556296  &   -74.560233&\\
 \hline
 F        & \textit{i}-FCIQMC$^b$      &      -99.55223(1)&       -99.64036(2)&      -99.68460(10)&      -99.70029(5)&\\
          & CIPSI                       &  -99.552228  &  -99.640295  &  -99.684561  &  -99.700258& -99.7341\\
          & CIPSI+LDA$_{\text{HF}}$     &   -99.658315  &   -99.704195  &   -99.722750  &   -99.727639\\
 \hline
 F$^+$    & \textit{i}-FCIQMC$^b$      &      -98.923015(6)&      -99.00542(1) &       -99.04599(3)&    -99.06082(4)&\\
          & CIPSI                       &  -98.923000  &  -99.005441  &  -99.046481  &  -99.060808& -99.0930\\
          & CIPSI+LDA$_{\text{HF}}$     &    -99.016909  &   -99.062981  &   -99.080847  &   -99.085872&\\
 \hline
 Ne       & \textit{i}-FCIQMC$^b$      &     -128.71145(3) &      -128.82577(5) &     -128.88065(6)&  -  &\\
          & CIPSI                       &  -128.711476  &  -128.825813  &  -128.880658  &  -128.900438 & -128.9383 \\
          & CIPSI+LDA$_{\text{HF}}$     &   -128.835474  &   -128.898894  &   -128.924219  &   -128.931038&\\
 \hline
 Ne$^+$   & \textit{i}-FCIQMC$^b$      &     -127.92411(2) &     -128.03691(2) &      -128.08816(11) & -              &   \\
          & CIPSI                       &  -127.924068  &  -128.036898  &  -128.088901  &  -128.107479& -128.1437\\
          & CIPSI+LDA$_{\text{HF}}$     &   -128.037019  &   -128.104203  &   -128.128973  &   -128.135914&
 \label{b-n-tot}
\end{tabular}
\end{ruledtabular}
\begin{flushleft}
\vspace{-0.2cm}
{$^a$ Estimated exact non-relativistic (NR) values from Ref. \onlinecite{exact_atoms}.}\\
{$^b$ From Ref. \onlinecite{ip_ali}. The statistical errors are given in parenthesis.}
\end{flushleft}
 \end{table*}

 \begin{table*}
 \caption{IPs (in mH) calculated by CIPSI and CIPSI+LDA$_\text{HF}$ for the B-Ne series with the AV$X$Z basis sets ($X=2, 3, 4, 5$). The errors with respect to the estimated exact non-relativistic values are given in parenthesis.}
 \begin{ruledtabular}
 \begin{tabular}{llddddd}
          &Method                    &      \multicolumn{1}{c}{   AV2Z}   &         \multicolumn{1}{c}{AV3Z}     &  \multicolumn{1}{c}{AV4Z}            &     \multicolumn{1}{c}{AV5Z}        & \multicolumn{1}{c}{Exact NR$^a$}\\
\cline{2-3} \cline{4-5} \cline{6-7}
 \hline
   B  & CIPSI                    &  297.92 $\;$ (7.05)   &  302.99  $\;$ (1.98)  &  304.06 $\;$ (0.91)   &  304.70  $\;$ (0.27)  &  304.98 \\   
      & CIPSI+LDA$_{\text{HF}}$  &  302.59 $\;$ (2.38)   &  305.12  $\;$ (-0.14) &  305.09 $\;$ (-0.11)  &  305.21  $\;$ (-0.23)      \\
\hline                                                                                                                    
   C  & CIPSI                    &  406.97 $\;$ (7.10)   &  411.92  $\;$ (2.15)  &  413.09 $\;$ (0.98)   &  413.67  $\;$ (0.40)  &  414.08 \\   
      & CIPSI+LDA$_{\text{HF}}$  &  411.64 $\;$ (2.43)   &  414.03  $\;$ (0.04)  &  414.14 $\;$ (-0.06)  &  414.22  $\;$ (-0.14)      \\
\hline                                                                                                                    
   N  & CIPSI                    &  527.75 $\;$ (7.13)   &  532.58  $\;$ (2.30)  &  533.82 $\;$ (1.06)   &  534.39  $\;$ (0.49)  &  534.89 \\   
      & CIPSI+LDA$_{\text{HF}}$  &  532.62 $\;$ (2.26)   &  534.75  $\;$ (0.13)  &  534.92 $\;$ (-0.03)  &  534.97  $\;$ (-0.08)      \\
\hline                                                                                                                    
   O  & CIPSI                    &  483.50 $\;$ (16.90)  &  493.73  $\;$ (6.67)  &  497.37 $\;$ (3.03)   &  498.89  $\;$ (1.51)  &  500.41 \\   
      & CIPSI+LDA$_{\text{HF}}$  &  497.29 $\;$ (3.11)   &  500.88  $\;$ (-0.47) &  501.59 $\;$ (-1.18)  &  501.40  $\;$ (-0.99)      \\
\hline                                                                                                                    
   F  & CIPSI                    &  629.22 $\;$ (11.90)  &  634.85  $\;$ (6.27)  &  638.07 $\;$ (3.05)   &  639.45  $\;$ (1.67)  &  641.13 \\   
      & CIPSI+LDA$_{\text{HF}}$  &  641.40 $\;$ (-0.27)  &  641.21  $\;$ (-0.08) &  641.90 $\;$ (-0.77)  &  641.76  $\;$ (-0.63)      \\
\hline                                                                                                                    
   Ne & CIPSI                    &  787.40 $\;$ (7.23)   &  788.91  $\;$ (5.72)  &  791.75 $\;$ (2.88)   &  792.95  $\;$ (1.68)  &  794.64\\
      & CIPSI+LDA$_{\text{HF}}$  &  798.45 $\;$ (-3.81)  &  794.69  $\;$ (-0.05) &  795.24 $\;$ (-0.60)  &  795.12  $\;$ (-0.48)
 \end{tabular}
 \label{b-ne-ip}
\end{ruledtabular}
\begin{flushleft}
\vspace{-0.3cm}
{$^a$ Estimated exact non-relativistic (NR) values from Ref. \onlinecite{exact_atoms}.}
\end{flushleft}
 \end{table*}

\subsubsection{The ionization potentials of the B-Ne series using CIPSI+LDA$_{\text{HF}}$}
\begin{figure*}[t]
\includegraphics[width=0.45\linewidth]{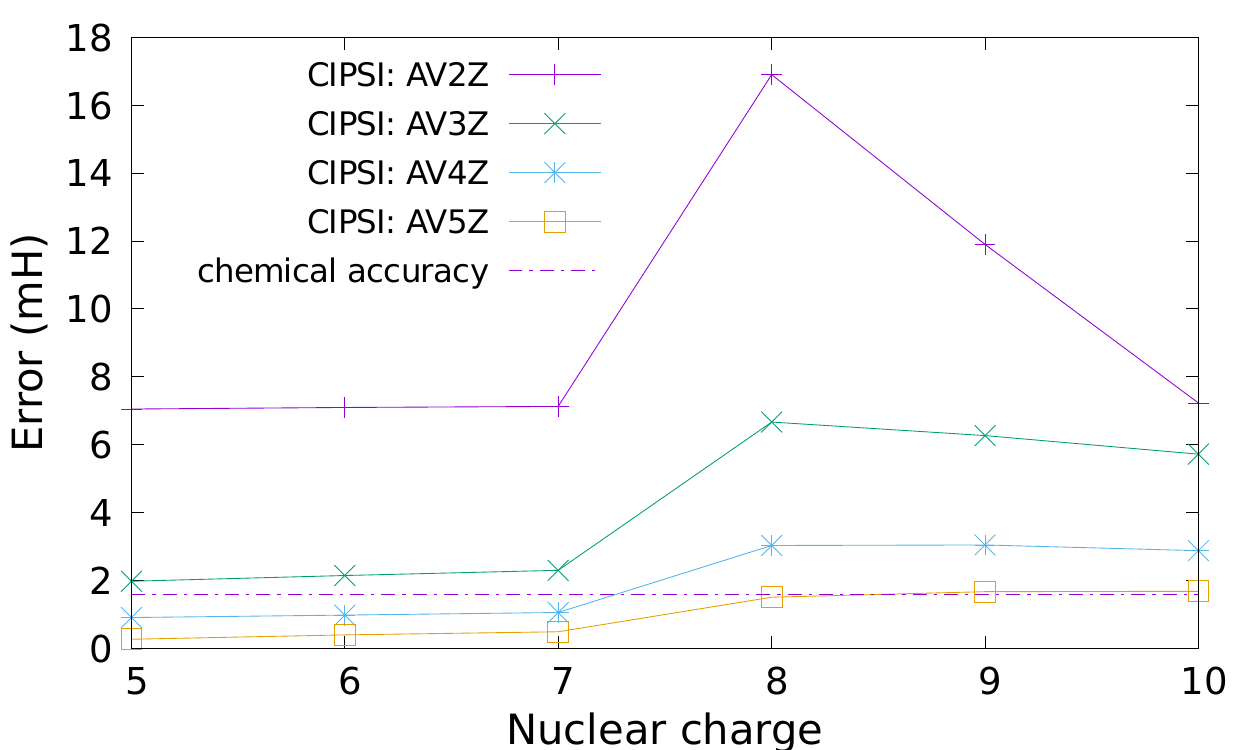}
\includegraphics[width=0.45\linewidth]{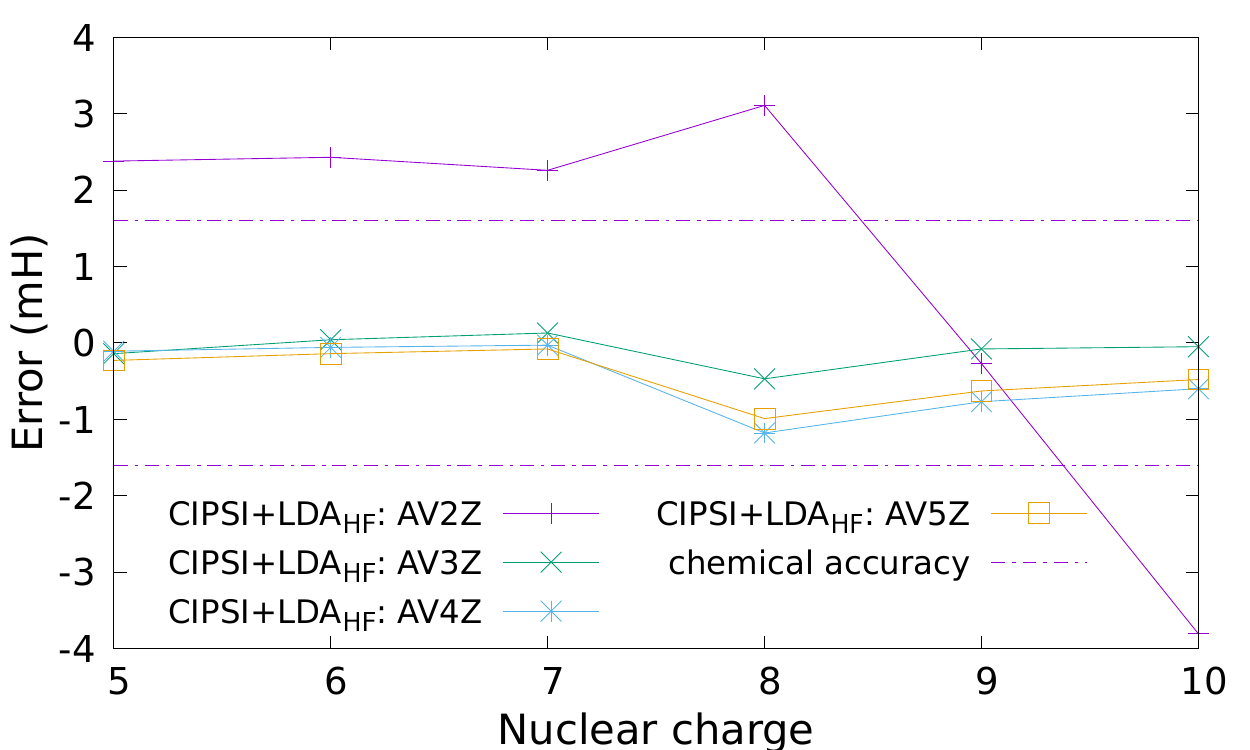}
  \caption{Errors on the IPs calculated at the CIPSI (left plot) and CIPSI+LDA$_{\text{HF}}$ (right plot) levels for the B-Ne series with the AV$X$Z basis sets ($X=2, 3, 4, 5$). Note the different scales of the two plots. }
\label{fig:err_ip_fci}
\end{figure*}

In order to investigate how the correction  $\efuncbasislda$ performs for energy differences, we report calculations of IPs for the B-Ne series using Dunning AV$X$Z basis sets ($X=2, 3, 4, 5$). These quantities have already been investigated at the initiator FCI quantum Monte Carlo (\textit{i}-FCIQMC) level by Alavi and coworkers\cite{ip_ali} and the authors have shown that obtaining errors of the IPs of the order of 1 mH for these simple atomic systems having at most ten electrons requires the use of large basis sets. As FCI in large basis sets is very rapidly out of reach for these systems, we use here the CIPSI+LDA$_{\text{HF}}$ method. The total energies are reported in table \ref{b-n-tot} and the IPs in table \ref{b-ne-ip}. A graphical representation of the errors with respect to the estimated exact non-relativistic IPs at the CIPSI and CIPSI+LDA$_{\text{HF}}$ levels is also reported in figure \ref{fig:err_ip_fci}.
All electrons were correlated in the CIPSI calculations and the calculations were stopped when $|E^{(2)}| < 10^{-3}$ Hartree, except for the Ne atom with the AV5Z basis set for which the calculation was stopped at $|E^{(2)}|=1.3\times10^{-3}$ Hartree.

From table \ref{b-n-tot} it clearly appears that all available \textit{i}-FCIQMC total energies values are perfectly reproduced by the CIPSI total energies, which can thus be considered as good approximations of the FCI energies. Also, considering the small threshold on $|E^{(2)}| $ and that the LDA correction $\efuncbasisldacipsihf$ converges very rapidly with respect to the number of Slater determinants (see figure \ref{fig:conv_cipsi}), the approximation of equation \eqref{eq:cipsidensity} can be considered as valid and therefore our CIPSI+LDA$_{\text{HF}}$ results can be considered as virtually identical to the ones that would be obtained with FCI+LDA$_{\text{HF}}$. Finally, the CIPSI+LDA$_{\text{HF}}$ total energies obtained with the AV5Z basis set are remarkably close to the estimated exact total energies for the whole series, with an error ranging from 3.8 mH for the B$^+$ cation to 7.2 mH for the Ne atom. 

Regarding the quality of the IPs (table \ref{b-ne-ip} and figure \ref{fig:err_ip_fci}), at near FCI level (either \textit{i}-FCIQMC or CIPSI) the typical chemical accuracy of 1 kcal/mol ($\approx$ 1.6 mH) is reached with the AV4Z basis set for the B, C, and N atoms, whereas such a level of accuracy is barely reached with the AV5Z basis set for the O, F, and Ne atoms. This illustrates how demanding the accurate computation of energy differences on these simple atomic systems is. Also one can notice that the IPs computed at the CIPSI level are systematically too small compared to the estimated exact values, showing that the cations are systematically better described than the neutral atoms in a given basis set. This result can be intuitively understood by the fact that the neutral atom has necessarily more correlated electron pairs than the cation and therefore, in the same basis, the cation is favored. 

Considering now the convergence of the results obtained at the CIPSI+LDA$_{\text{HF}}$ level with respect to the basis set, it is striking to observe how the addition of the DFT correction improves the accuracy of the energy differences, with a sub kcal/mol accuracy being obtained for all atoms from the AV3Z to the AV5Z basis sets. With the AV2Z basis set, the error is overall strongly reduced, the average error being about 3 mH at the CIPSI+LDA$_{\text{HF}}$, whereas it is of about 9 mH at the CIPSI level. From the AV3Z and larger basis sets, the maximum error occurs for the IP of the oxygen atom, which is overestimated by only 1.1 mH with the AV4Z basis set and by 0.9 mH with the AV5Z basis set, showing the accuracy of the approach. One can nevertheless observe a global trend of CIPSI+LDA$_{\text{HF}}$ to overestimate the IP, which is due to an over-correlation of the neutral species. 

\begin{figure*}[t]
\includegraphics[width=0.45\linewidth]{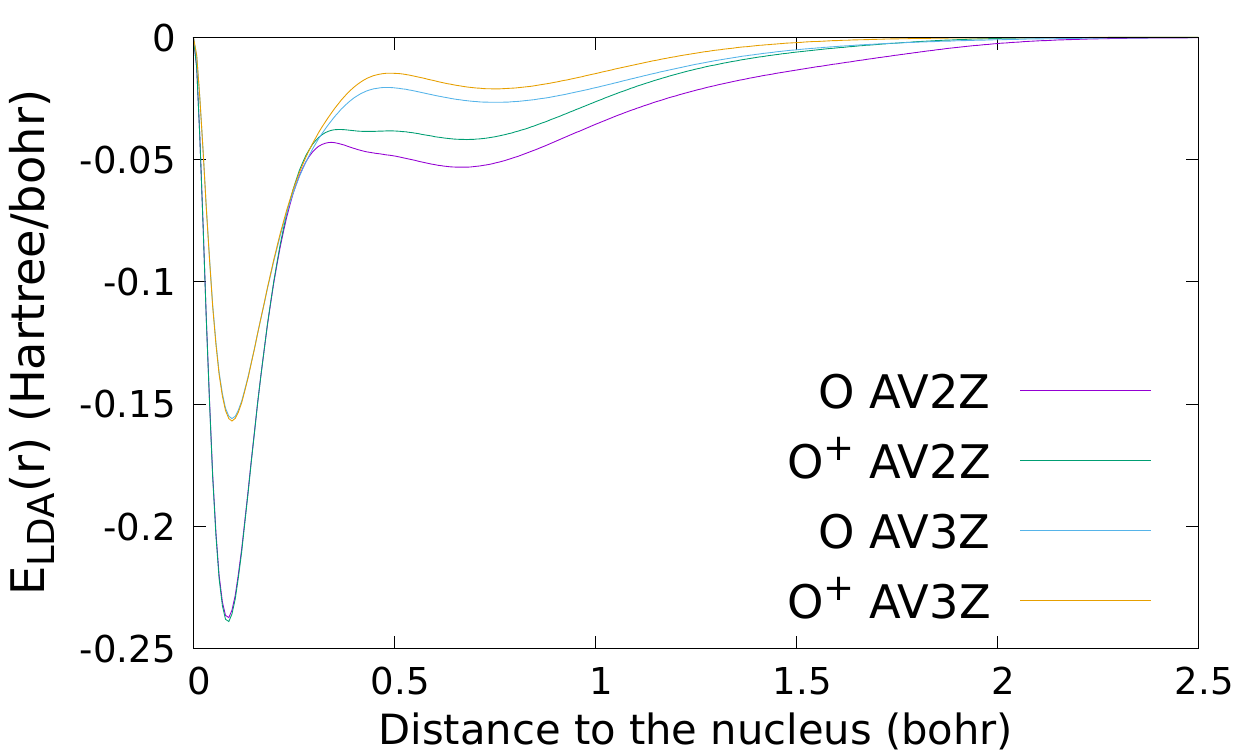}
\includegraphics[width=0.45\linewidth]{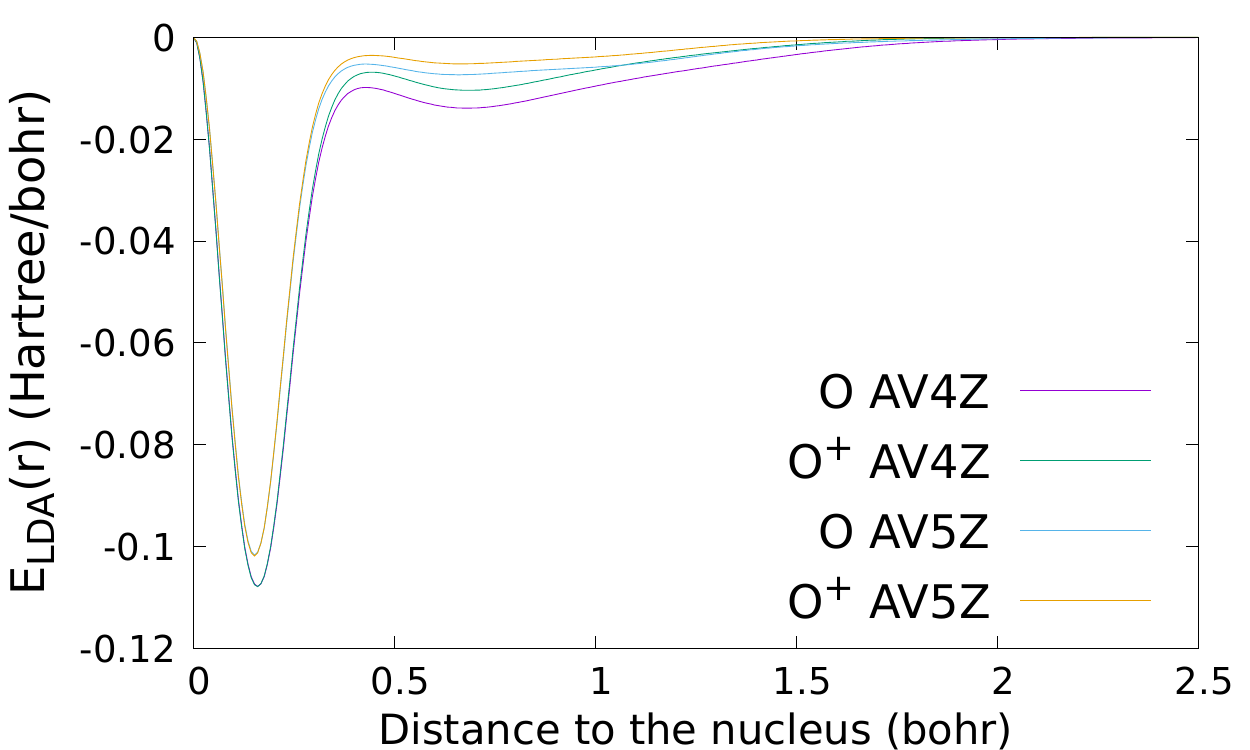}
  \caption{Behavior of $\bar{E}_{\text{LDA}}^{\basis}(r)$ for the AV2Z and AV3Z basis sets (left plot) and AV4Z and AV5Z basis sets (right plot) for the oxygen atom and its first cation.}
\label{fig:ec_of_r_dz}
\end{figure*}

\subsubsection{A case study: The oxygen atom and cation}
\label{sec:numerical_o}
In order to better understand how $\efuncbasisldacipsihf$ corrects for the basis-set incompleteness and its impact on the energy differences, we perform a detailed study of the behavior of two quantities related to $\efuncbasisldacipsihf$ for the oxygen atom and its first cation. 

We first define the spherically averaged local basis-set correction as
\begin{equation}
 \begin{aligned}
 \bar{E}_{\text{LDA}}^{\basis}(r) = \iint  \text{d}\Omega \,\, r^2  &\dencipsi 
\nonumber\\
& \,\,\emuldahf ,
 \end{aligned}
\end{equation}
such that 
\begin{equation}
  \int \text{d}r\,\,\bar{E}_{\text{LDA}}^{\basis}(r) = \efuncbasisldacipsihf \, ,
\end{equation}
where we use the largest CIPSI wave function to obtain the density $\dencipsi$. 
With $\bar{E}_{\text{LDA}}^{\basis}(r)$ one can analyze in real space how $\efuncbasisldacipsihf$ corrects for the incompleteness of the basis set in near FCI calculations. 


We report in figure \ref{fig:ec_of_r_dz} the plot of $\bar{E}_{\text{LDA}}^{\basis}(r)$ for the oxygen atom and its cation for different basis sets. One can observe that, with all basis sets used here, the LDA correction for the neutral atom is overall larger in absolute value than for the cation, which confirms that the cation is better described in a given basis set than the neutral atom. Also, it clearly explains why $\efuncbasisldacipsihf$ has a differential effect on the IPs. Regarding the behavior as a function of the distance to the nucleus, all the curves show that the dominant contributions, in absolute value, are in the region of high density. As expected, $\bar{E}_{\text{LDA}}^{\basis}(r)$ gets smaller as the size of the basis set is increased. With the largest basis set, $\bar{E}_{\text{LDA}}^{\basis}(r)$ is small in the valence shell ($r>0.5$ bohr), but remains substantial in the core region. The fact that the basis sets used here do not contain functions optimized for core correlation explains why the LDA correction remains important in the core region, even with the AV5Z basis set.

\begin{figure*}
\includegraphics[width=0.45\linewidth]{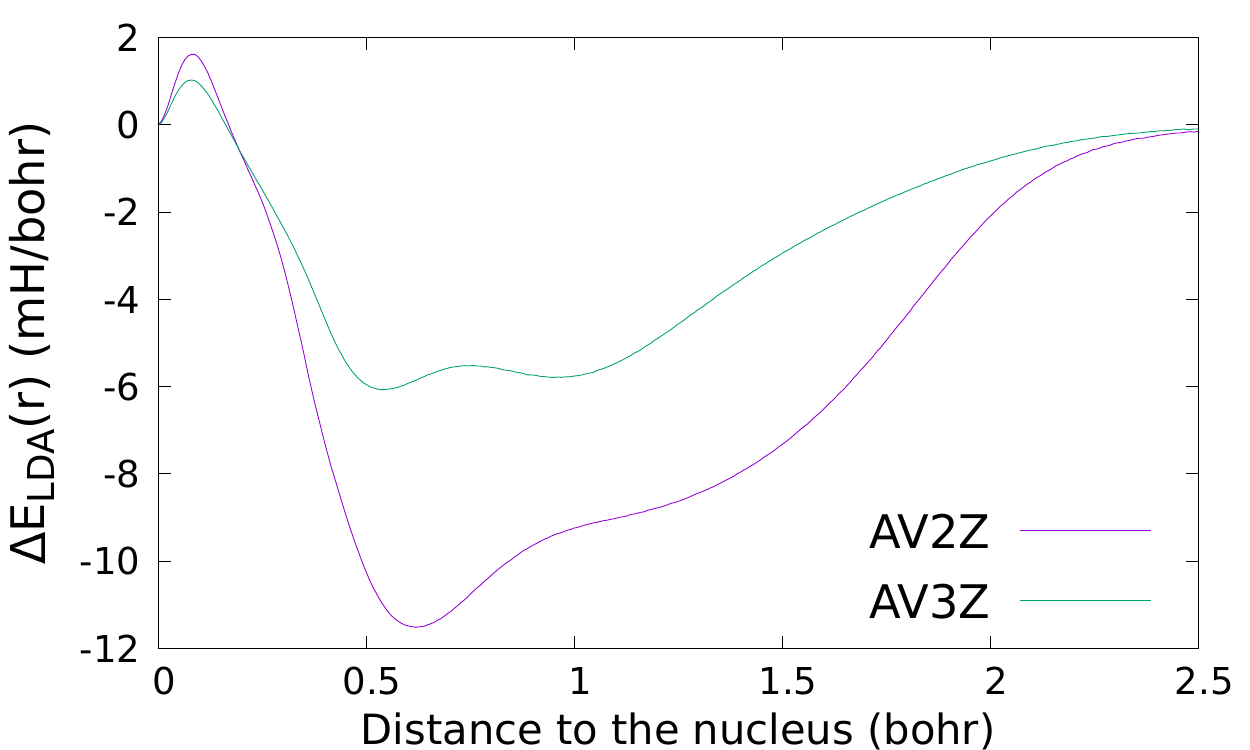}
\includegraphics[width=0.45\linewidth]{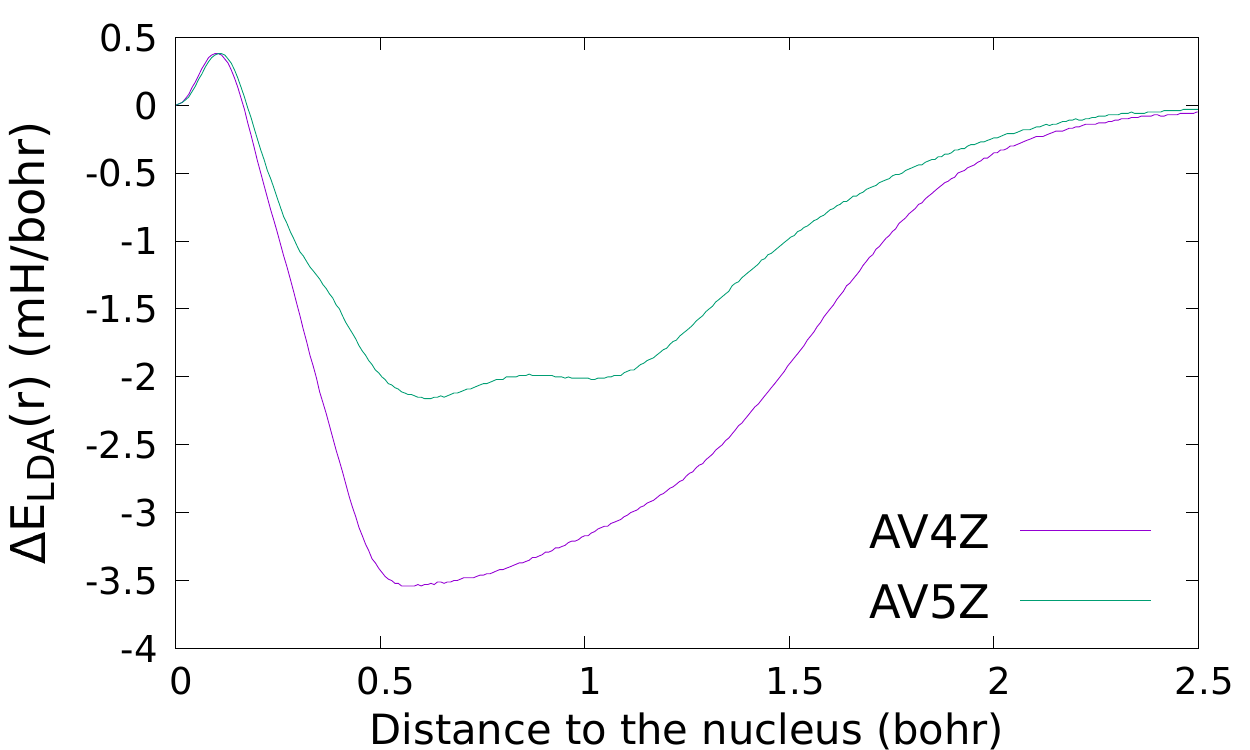}
 \caption{Behavior of $\Delta \bar{E}_{\text{LDA}}^{\basis}(r)$ for the AV2Z and AV3Z basis sets (left plot) and AV4Z and AV5Z basis sets (right plot) for the oxygen atom.}
\label{fig:delta_ec_of_r_dz}
\end{figure*}

In order to investigate the differential impact of the DFT correction on O and O$^+$, we also define the following function: 
\begin{equation}
 \Delta \bar{E}_{\text{LDA}}^{\basis}(r) = \bar{E}_{\text{LDA},\text{O}}^{\basis}(r) - \bar{E}_{\text{LDA},\text{O}^+}^{\basis}(r) \, .
\end{equation}
We report in figure \ref{fig:delta_ec_of_r_dz} the values of $\Delta \bar{E}_{\text{LDA}}^{\basis}(r) $ for different basis sets. It clearly appears that the differential effects are mainly located in the valence region, which is what is expected since the electron can be qualitatively considered to be removed from the valence region. Also, except for the inner core region, $\Delta \bar{E}_{\text{LDA}}^{\basis}(r) $ is always negative which means that $\efuncbasisldacipsihf$ corrects more the neutral atom than the cation for the basis-set incompleteness. The fact that $\Delta \bar{E}_{\text{LDA}}^{\basis}(r) $ is positive near 0.1 bohr means that the cation is more correlated in this region, which could be a sign that the two 1s electrons are closer to each other in the cation than in the neutral atom.

\section{Conclusion}
\label{sec:conclusion}
In the present work we have proposed a theory based on DFT to correct for the basis-set incompleteness of WFT. The key point here is the definition of a local range-separation parameter $\mu({\bf r})$ which automatically adapts to the basis set. 

Both the exact theory (see section \ref{sec:operator_basis}) and a series of approximations (see sections \ref{sec:fci_density}, \ref{sec:practical_approx}, and \ref{sec:cipsi_lda}) were derived for FCI and selected CI wave functions. Our theory combines WFT with a complementary density functional, as in RS-DFT. Unlike the latter theory, the electron-electron interaction is split directly in the one-electron basis set (see section \ref{sec:operator_basis}). Here, the part of the electron-electron interaction expanded in the basis set is treated by WFT and the remaining interaction by the density functional. Thanks to a definition of the real-space representation of the basis-set-projected electron-electron interaction (see section \ref{sec:real_Coulomb}), we show that the effect of the incompleteness of a given basis set can be mapped into a non-diverging effective electron-electron interaction. We derive some of the important exact properties of the effective electron-electron interaction (see Appendix \ref{sec:complete_basis} and section \ref{sec:trncated_basis}), which helps us to physically motivate such a choice for an effective electron-electron interaction. A mapping between RS-DFT and our theory is proposed through the non-diverging behavior of the interactions in both theories (see section \ref{sec:qualitative}), and such a mapping is done in practice through a comparison at coalescence of the effective electron-electron interaction with the long-range interaction used in the RS-DFT framework (see section \ref{sec:mu_of_r}). More specifically, this link between the basis-set splitting and range separation of the electron-electron interaction is done through the definition of a range-separation parameter $\mu({\bf r})$ which now depends on the spatial coordinate in $\real{3}$. The computation of $\mu({\bf r})$  nonetheless requires the computation of bi-electronic integrals. This allows us to benefit from all pre-existing methodologies developed in the RS-DFT framework and therefore to produce numerically tractable approximations for our theory (see section \ref{sec:practical_approx} for the definition of an LDA-like functional in the present context). As the local range-separation parameter $\mu({\bf r})$ is automatically defined for a given physical system in a given basis set, we completely remove the choice of the parameter $\mu$ which is inherent in the RS-DFT framework. Also this local range-separation parameter $\mu({\bf r})$  can be seen as a measure of the incompleteness of a given basis set together with its non uniformity in the description of the correlation effects in $\real{3}$. Finally, our theory produces a DFT-based correction for a given basis set which is added to the approximation of the FCI energy obtained in the same basis set. 

We performed numerical tests both for total energies and energy differences for atomic systems (see section \ref{sec:numerical}). Using FCI wave functions (see section \ref{sec:fci_numerical}), we demonstrated that our approach is able to accelerate the basis convergence toward the exact non-relativistic total energy for the helium atom, which numerically illustrates its systematically improvable character. Then, we investigated the accuracy of our basis-set corrected CIPSI approach to describe the IPs of the B-Ne series (see section \ref{sec:cipsi_numerical}) which are known to be challenging for WFT methods, even at near FCI level. The main result of this study is that the level of accuracy of the energy differences is drastically improved even using the small aug-cc-pVDZ basis set, and that a sub-kcal/mol error is reached for all atoms from the aug-cc-pVTZ up to the aug-cc-pV5Z basis sets. Such results have to be compared with near FCI results for which a comparable error is barely reached only using the aug-cc-pV5Z basis set. In order to have a better understanding of the origin of the systematic improvement of IPs brought by the DFT correction, we performed a detailed study of the oxygen atom and its first cation (see section \ref{sec:numerical_o}). By introducing spherical averaged quantities, we show that the major differential contribution brought by the DFT correction comes from the valence region, which is physically meaningful and therefore tends to confirm that the good results obtained with our approach do not come from fortuitous error cancellations. 
Finally, it is important to stress here that the computational cost of the DFT corrections used here represents a negligible percentage of the computational cost of the CIPSI calculations.

\appendix
\section{Derivation of the real-space representation of the effective interaction projected in a basis set}
\label{sec:appendix_expectation}
The exact Coulomb electron-electron operator can be expressed in real-space second quantization as
\begin{equation}
\label{eq:Weerealspace}
   \begin{aligned}
     \weeop =  \frac{1}{2}\,\,\iiiint  &\dr{1}\,\dr{2} \,\dr{3}\,\dr{4} \\ & \delta (\bfr{1}-\bfr{4})  \,\,\delta (\bfr{2}-\bfr{3})  \frac{1}{\norm{\bfrb{1} - \bfrb{2} }} \\
              & \psixc{4}\psixc{3} \psix{2}\psix{1},
   \end{aligned}
\end{equation}
where $\psix{}$ and $\psixc{}$ are annihilation and creation field operators, and $\bfr{}=\left({\bf r},\sigma \right)$ collects the space and spin variables. The Coulomb electron-electron operator restricted to a basis set $\basis$ can be written in orbital-space second quantization:
\begin{equation}
   \begin{aligned}
     \weeopbasis =  \frac{1}{2}\,\, \sum_{ijkl\,\,\in\,\,\basis} \,\, \vijkl \,\, \aic{k}\aic{l}\ai{j}\ai{i},
   \end{aligned}
\end{equation}
where the summations run over all (real-valued) orthonormal spin-orbitals $\{\phix{i}{} \}$ in the basis set $\basis$, $\vijkl$ are the two-electron integrals, the annihilation and creation operators can be written in terms of the field operators as
\begin{equation}
  \ai{i} = \int \dr{} \,\, \phix{i}{}\,\,\psix{},
\end{equation}
\begin{equation}
  \aic{i} = \int \dr{} \,\, \phix{i}{}\,\,\psixc{}.
\end{equation}
Therefore, by defining 
\begin{equation}
   \begin{aligned}
    & w^\basis(\bfr{1},\bfr{2},\bfr{3},\bfr{4})  \\= & \sum_{ijkl\,\,\in\,\,\basis}   \,\, \vijkl \,\,  \phix{k}{4} \phix{l}{3} \phix{j}{2} \phix{i}{1}\, ,
   \end{aligned}
\end{equation}
we can rewrite $ \weeopbasis$ in real-space second quantization as
\begin{equation}
\label{eq:WeeBrealspace}
   \begin{aligned}
     \weeopbasis =  \frac{1}{2}\,\,&\iiiint \dr{1}\,\dr{2} \,\dr{3}\,\dr{4} \\ & w^\basis(\bfr{1},\bfr{2},\bfr{3},\bfr{4}) \\ & \psixc{4}\psixc{3} \psix{2}\psix{1}.
   \end{aligned}
\end{equation}
In the limit of a complete basis set (written as ``$\basis \rightarrow \infty$''), $\weeopbasis$ coincides with $\weeop$: 
\begin{equation}
  \begin{aligned}
    \lim_{\basis \rightarrow \infty}  \weeopbasis = \weeop,
  \end{aligned}
\end{equation}
which implies that
\begin{equation}
  \begin{aligned}
    & \lim_{\basis \rightarrow \infty}   w^\basis(\bfr{1},\bfr{2},\bfr{3},\bfr{4}) =\\&  \delta (\bfr{1}-\bfr{4})  \,\,\delta (\bfr{2}-\bfr{3}) \,\,\frac{1}{\norm{\bfrb{1} - \bfrb{2}} }.
  \end{aligned}
\end{equation}
It is important here to stress that the definition $ w^\basis(\bfr{1},\bfr{2},\bfr{3},\bfr{4})$ tends to a distribution in the limit of a complete basis set, and therefore such an object must really be considered as a distribution acting on test functions and not as a function to be evaluated pointwise. This is why we need to use an expectation value in order to make sense out of $ w^\basis(\bfr{1},\bfr{2},\bfr{3},\bfr{4})$.

From equation \eqref{eq:Weerealspace}, the expectation value of the Coulomb electron-electron operator over a wave function $\Psi$ is, after integration over $\bfr{3}$ and $\bfr{4}$, 
\begin{equation}
 \begin{aligned}
   \elemm{\Psi}{\weeop}{\Psi} =  &\frac{1}{2}\,\,\iint  \dr{1}\,\dr{2} \,\,\frac{1}{\norm{\bfrb{1} - \bfrb{2}}} \\ &\elemm{\Psi}{\psixc{1}\psixc{2} \psix{2}\psix{1}}{\Psi},
 \end{aligned}
\end{equation}
which, by introducing the two-body density matrix,
\begin{equation}
  \begin{aligned}
    & \twodmr{1}{2}{3}{4}{\Psi} \\ & = \twodm{1}{2}{3}{4},
  \end{aligned}
\end{equation}
turns into 
\begin{equation}
\label{eqapp:PsiWPsi}
  \begin{aligned}
    \elemm{\Psi}{\weeop}{\Psi} =  \frac{1}{2}\,\,\iint  \dr{1}\,\dr{2} \,\,\frac{1}{\norm{\bfrb{1} - \bfrb{2} }} \,\,  \twodmrdiag{1}{2}{\Psi},
  \end{aligned}
\end{equation}
where $\twodmrdiag{1}{2}{\Psi}=\twodmr{1}{2}{2}{1}{\Psi}$ is the pair density of $\Psi$. Equation \eqref{eqapp:PsiWPsi} holds for any wave function $\Psi$. Consider now the expectation value of $\weeopbasis$ over a wave function $\psibasis$. From equation \eqref{eq:WeeBrealspace}, we get
\begin{equation}
   \label{eq:wbasis_elemm}
   \begin{aligned}
     \elemm{\psibasis}{\weeopbasis}{\psibasis} =  &\,\,\frac{1}{2}\,\,\iiiint  \dr{1}\,\dr{2} \,\dr{3}\,\dr{4} \\ &w^\basis(\bfr{1},\bfr{2},\bfr{3},\bfr{4})  \twodmr{1}{2}{3}{4}{\psibasis},
   \end{aligned}
\end{equation}
where $\twodmr{1}{2}{3}{4}{\psibasis}$ is expressed as 
\begin{equation}
 \begin{aligned}
  &\twodmr{1}{2}{3}{4}{\psibasis} =\\ & \sum_{mnpq\,\,\in\,\,\basis} \phix{p}{4} \phix{q}{3} \phix{n}{2} \phix{m}{1} \,\, \gammamnpq{\psibasis},
 \end{aligned}
\end{equation}
and $\gammamnpq{\psibasis}$ is the two-body density tensor of $\psibasis$
\begin{equation}
  \gammamnpq{\psibasis} = \elemm{\psibasis}{ \aic{p}\aic{q}\ai{n}\ai{m} }{\psibasis}.
\end{equation}
By integrating over $\bfr{3}$ and $\bfr{4}$ in equation \eqref{eq:wbasis_elemm}, it comes:
\begin{equation}
  \elemm{\psibasis}{\weeopbasis}{\psibasis} =  \frac{1}{2}\,\,\iint \dr{1}\,\dr{2} \,\, \fbasis,
\end{equation}
where we introduced the function
\begin{equation}
  \label{eqapp:fbasis}
  \begin{aligned}
  \fbasis =  \sum_{ijklmn\,\,\in\,\,\basis} & \vijkl \,\, \gammaklmn{\psibasis} \\& \phix{n}{2} \phix{m}{1}  \phix{i}{1} \phix{j}{2}.
  \end{aligned}
\end{equation}
From the definition of the restriction of an operator to the space generated by the basis set $\basis$, we have the following equality
\begin{equation}
  \elemm{\psibasis}{\weeopbasis}{\psibasis} = \elemm{\psibasis}{\weeop}{\psibasis},
\end{equation}
which translates into
\begin{equation}
  \begin{aligned}
    \label{eqapp:int_eq_basis}
    & \frac{1}{2}\,\,\iint  \dr{1}\,\dr{2} \,\,\fbasis \\ =\,\,&\frac{1}{2}\,\,\iint \dr{1}\,\dr{2} \,\,\frac{1}{\norm{\bfrb{1} - \bfrb{2} }} \,\,  \twodmrdiag{1}{2}{\psibasis},
  \end{aligned}
\end{equation}
and holds for any $\psibasis$.
Therefore, by introducing the following function 
\begin{equation}
  \label{eqapp:def_weebasis}
  \wbasis = \frac{\fbasis}{\twodmrdiagpsi},
\end{equation}
one can rewrite equation \eqref{eqapp:int_eq_basis} as
\begin{equation}
  \begin{aligned}
    \label{eqapp:int_eq_wee}
    & \iint \dr{1}\,\dr{2} \,\, \wbasis \,\, \twodmrdiagpsi \\ = &\iint \dr{1}\,\dr{2} \,\,\frac{1}{\norm{\bfrb{1} - \bfrb{2} }} \,\,  \twodmrdiagpsi.
  \end{aligned}
\end{equation}

\section{Behavior of the effective electron-electron interaction $\wbasis$ in the limit of a complete basis set}
\label{sec:complete_basis}
To study how $\wbasis$ behaves in the limit of a complete basis set, one only needs to study $\fbasis$. 
By expliciting the two-electron integrals, $\fbasis$ can be written as
\begin{equation}
  \begin{aligned}
    & \fbasis  =  \\ & \sum_{ijklmn\,\in\,\basis}    \gammaklmn{\psibasis}\phix{n}{2} \phix{m}{1}  \phix{i}{1} \phix{j}{2}  \\ &\iint \dr{}\,\dr{}' \,\, \phix{k}{} \phixprim{l}{}  \phix{i}{}  \phixprim{j}{} \,\,\frac{1}{\norm{\bfrb{} - \bfrb{}'}},
  \end{aligned}
\end{equation}
which, after regrouping the summations over the indices $i$ and $j$, becomes:
\begin{equation}
    \label{eq:f_b}
  \begin{aligned}
    &\fbasis =  \sum_{mnkl\,\,\in\,\,\basis}  \gammaklmn{\psibasis} \phix{n}{2} \phix{m}{1}  \\
    & \int \dr{} \,\,  \left( \,\sum_{i\,\,\in\,\,\basis} \phix{i}{1}  \phix{i}{}\right) \phix{k}{} \\ & \int \,\dr{}'\,\,\left( \,\sum_{j\,\,\in\,\,\basis} \phix{j}{2} \phixprim{j}{} \right) \phixprim{l}{} \,\,\frac{1}{\norm{\bfrb{} - \bfrb{}'}}.
  \end{aligned}
\end{equation}
One can recognize in equation \eqref{eq:f_b} the expression of the restriction of a Dirac distribution to the basis set $\basis$:
\begin{equation}
  \delta^{\basis} ({\bf Y} - {\bf Y}') = \sum_{i\,\,\in\,\,\basis}\,\,\phi_i({\bf Y}) \phi_i({\bf Y}').
\end{equation}
Such a distribution  $\delta^{\basis} ({\bf Y} - {\bf Y}')$  maintains the standard Dirac distribution properties only when applied to functions which are exactly representable in $\basis$. 
More precisely, if $g$ is a test function from $\mathbb{R}^3$ to $\mathbb{R}$, $g^\basis$ its component in $\basis$ and $g^\perp$ the orthogonal component:
\begin{equation}
   g = g^\basis + g^\perp \;\; \text{with} \;\;
  \int \text{d}{\bf r} \,\,  g^\basis({\bf r}) \,\, g^\perp({\bf r}) = 0,
\end{equation}
then 
\begin{equation}
  \int \text{d}{\bf r} \,\,  \delta^{\basis}({\bf r} - {\bf r}') \,\, g({\bf r}) \,\, =  \,\, g({\bf r}') \,\,\, \text{iff} \,\, g^\perp ({\bf r}')= 0 \,\,\,\, \forall \,\, {\bf r}'.
\end{equation}
In the limit of a complete basis set, the function $\phixprim{l}{} \,\,\frac{1}{\norm{\bfrb{} - \bfrb{}'}}$ is necessarily within $\basis$, and thus one has:
\begin{equation}
  \begin{aligned}
    \lim_{\basis \rightarrow \infty} & \int \,\dr{}'\,\, \delta^{\basis} ({\bf X}_2 - {\bf X}') \phixprim{l}{} \,\,\frac{1}{\left|\bfrb{} - \bfrb{}'\right    |} \\ &= \phix{l}{2} \,\,\frac{1}{\norm{\bfrb{} - \bfrb{2}}},
  \end{aligned}
\end{equation}
and
\begin{equation}
  \begin{aligned}
    \lim_{\basis \rightarrow \infty}   \int \dr{} \,\, \delta^{\basis} ({\bf X}_1 - {\bf X})  \phix{k}{} & \int \,\dr{}'\,\,\frac{ \delta^{\basis} ({\bf X}_2 - {\bf X}')   \phixprim{l}{} }{\norm{\bfrb{} - \bfrb{}'}}\\
     = &\,\,\phix{k}{1} \,\,  \phix{l}{2} \,\,\frac{1}{\norm{\bfrb{1} - \bfrb{2}}}.
  \end{aligned}
\end{equation}
Inserting this expression in $\fbasis$ leads to:
\begin{equation}
 \begin{aligned}
   & \lim_{\basis \rightarrow \infty}  \fbasis  = \sum_{klmn\,\,\in\,\,\basis}  \,\, \gammaklmn{\psibasis} \\  & \phix{m}{1} \,\, \phix{n}{2} \,\,  \phix{l}{2} \,\,\phix{k}{1} \,\,\frac{1}{\norm{\bfrb{1} - \bfrb{2}}},
 \end{aligned}
\end{equation}
which is nothing but
\begin{equation}
  \lim_{\basis \rightarrow \infty}  \fbasis  =  \twodmrdiagpsi \,\,\frac{1}{\norm{\bfrb{1} - \bfrb{2}}}.
\end{equation}
Therefore, in the limit of a complete basis set, the effective electron-electron interaction $\wbasis$ correctly reduces to the true Coulomb interaction interaction for all points $(\bfr{1},\bfr{2})$.
\begin{equation}
   \label{eqapp:weeb_infty}
    \lim_{\basis \rightarrow \infty} \wbasis  = \frac{1}{\norm{\bfrb{1} - \bfrb{2}}}, \quad \forall \,\, (\bfr{1},\bfr{2}) \,\, \text{and }\psibasis.
\end{equation}


%

\begin{thebibliography}{0}%
\makeatletter
\providecommand \@ifxundefined [1]{%
 \@ifx{#1\undefined}
}%
\providecommand \@ifnum [1]{%
 \ifnum #1\expandafter \@firstoftwo
 \else \expandafter \@secondoftwo
 \fi
}%
\providecommand \@ifx [1]{%
 \ifx #1\expandafter \@firstoftwo
 \else \expandafter \@secondoftwo
 \fi
}%
\providecommand \natexlab [1]{#1}%
\providecommand \enquote  [1]{``#1''}%
\providecommand \bibnamefont  [1]{#1}%
\providecommand \bibfnamefont [1]{#1}%
\providecommand \citenamefont [1]{#1}%
\providecommand \href@noop [0]{\@secondoftwo}%
\providecommand \href [0]{\begingroup \@sanitize@url \@href}%
\providecommand \@href[1]{\@@startlink{#1}\@@href}%
\providecommand \@@href[1]{\endgroup#1\@@endlink}%
\providecommand \@sanitize@url [0]{\catcode `\\12\catcode `\$12\catcode
  `\&12\catcode `\#12\catcode `\^12\catcode `\_12\catcode `\%12\relax}%
\providecommand \@@startlink[1]{}%
\providecommand \@@endlink[0]{}%
\providecommand \url  [0]{\begingroup\@sanitize@url \@url }%
\providecommand \@url [1]{\endgroup\@href {#1}{\urlprefix }}%
\providecommand \urlprefix  [0]{URL }%
\providecommand \Eprint [0]{\href }%
\providecommand \doibase [0]{http://dx.doi.org/}%
\providecommand \selectlanguage [0]{\@gobble}%
\providecommand \bibinfo  [0]{\@secondoftwo}%
\providecommand \bibfield  [0]{\@secondoftwo}%
\providecommand \translation [1]{[#1]}%
\providecommand \BibitemOpen [0]{}%
\providecommand \bibitemStop [0]{}%
\providecommand \bibitemNoStop [0]{.\EOS\space}%
\providecommand \EOS [0]{\spacefactor3000\relax}%
\providecommand \BibitemShut  [1]{\csname bibitem#1\endcsname}%
\let\auto@bib@innerbib\@empty
\end{thebibliography}%


\begin{thebibliography}{50}%
\makeatletter
\providecommand \@ifxundefined [1]{%
 \@ifx{#1\undefined}
}%
\providecommand \@ifnum [1]{%
 \ifnum #1\expandafter \@firstoftwo
 \else \expandafter \@secondoftwo
 \fi
}%
\providecommand \@ifx [1]{%
 \ifx #1\expandafter \@firstoftwo
 \else \expandafter \@secondoftwo
 \fi
}%
\providecommand \natexlab [1]{#1}%
\providecommand \enquote  [1]{``#1''}%
\providecommand \bibnamefont  [1]{#1}%
\providecommand \bibfnamefont [1]{#1}%
\providecommand \citenamefont [1]{#1}%
\providecommand \href@noop [0]{\@secondoftwo}%
\providecommand \href [0]{\begingroup \@sanitize@url \@href}%
\providecommand \@href[1]{\@@startlink{#1}\@@href}%
\providecommand \@@href[1]{\endgroup#1\@@endlink}%
\providecommand \@sanitize@url [0]{\catcode `\\12\catcode `\$12\catcode
  `\&12\catcode `\#12\catcode `\^12\catcode `\_12\catcode `\%12\relax}%
\providecommand \@@startlink[1]{}%
\providecommand \@@endlink[0]{}%
\providecommand \url  [0]{\begingroup\@sanitize@url \@url }%
\providecommand \@url [1]{\endgroup\@href {#1}{\urlprefix }}%
\providecommand \urlprefix  [0]{URL }%
\providecommand \Eprint [0]{\href }%
\providecommand \doibase [0]{http://dx.doi.org/}%
\providecommand \selectlanguage [0]{\@gobble}%
\providecommand \bibinfo  [0]{\@secondoftwo}%
\providecommand \bibfield  [0]{\@secondoftwo}%
\providecommand \translation [1]{[#1]}%
\providecommand \BibitemOpen [0]{}%
\providecommand \bibitemStop [0]{}%
\providecommand \bibitemNoStop [0]{.\EOS\space}%
\providecommand \EOS [0]{\spacefactor3000\relax}%
\providecommand \BibitemShut  [1]{\csname bibitem#1\endcsname}%
\let\auto@bib@innerbib\@empty
\bibitem [{\citenamefont {Goldstone}(1957)}]{goldstone}%
  \BibitemOpen
  \bibfield  {author} {\bibinfo {author} {\bibfnamefont {J.}~\bibnamefont
  {Goldstone}},\ }\href {\doibase 10.1098/rspa.1957.0037} {\bibfield  {journal}
  {\bibinfo  {journal} {Proc. R. Soc. A}\ }\textbf {\bibinfo {volume} {239}},\
  \bibinfo {pages} {267} (\bibinfo {year} {1957})}\BibitemShut {NoStop}%
\bibitem [{\citenamefont {Lindgren}(1985)}]{lindgren}%
  \BibitemOpen
  \bibfield  {author} {\bibinfo {author} {\bibfnamefont {I.}~\bibnamefont
  {Lindgren}},\ }\href {http://stacks.iop.org/1402-4896/32/i=4/a=009}
  {\bibfield  {journal} {\bibinfo  {journal} {Phys. Scr.}\ }\textbf {\bibinfo
  {volume} {32}},\ \bibinfo {pages} {291} (\bibinfo {year} {1985})}\BibitemShut
  {NoStop}%
\bibitem [{\citenamefont {Bartlett}\ and\ \citenamefont
  {Musia\l{}}(2007)}]{review_cc_bartlett}%
  \BibitemOpen
  \bibfield  {author} {\bibinfo {author} {\bibfnamefont {R.~J.}\ \bibnamefont
  {Bartlett}}\ and\ \bibinfo {author} {\bibfnamefont {M.}~\bibnamefont
  {Musia\l{}}},\ }\href {\doibase 10.1103/RevModPhys.79.291} {\bibfield
  {journal} {\bibinfo  {journal} {Rev. Mod. Phys.}\ }\textbf {\bibinfo {volume}
  {79}},\ \bibinfo {pages} {291} (\bibinfo {year} {2007})}\BibitemShut
  {NoStop}%
\bibitem [{\citenamefont {Bender}\ and\ \citenamefont
  {Davidson}(1969)}]{bender}%
  \BibitemOpen
  \bibfield  {author} {\bibinfo {author} {\bibfnamefont {C.~F.}\ \bibnamefont
  {Bender}}\ and\ \bibinfo {author} {\bibfnamefont {E.~R.}\ \bibnamefont
  {Davidson}},\ }\href {\doibase 10.1103/PhysRev.183.23} {\bibfield  {journal}
  {\bibinfo  {journal} {Phys. Rev.}\ }\textbf {\bibinfo {volume} {183}},\
  \bibinfo {pages} {23} (\bibinfo {year} {1969})}\BibitemShut {NoStop}%
\bibitem [{\citenamefont {Huron}, \citenamefont {Malrieu},\ and\ \citenamefont
  {Rancurel}(1973)}]{malrieu}%
  \BibitemOpen
  \bibfield  {author} {\bibinfo {author} {\bibfnamefont {B.}~\bibnamefont
  {Huron}}, \bibinfo {author} {\bibfnamefont {J.}~\bibnamefont {Malrieu}}, \
  and\ \bibinfo {author} {\bibfnamefont {P.}~\bibnamefont {Rancurel}},\ }\href
  {\doibase 10.1063/1.1679199} {\bibfield  {journal} {\bibinfo  {journal} {J.
  Chem. Phys.}\ }\textbf {\bibinfo {volume} {58}},\ \bibinfo {pages} {5745}
  (\bibinfo {year} {1973})}\BibitemShut {NoStop}%
\bibitem [{\citenamefont {Buenker}\ and\ \citenamefont
  {Peyerimholf}(1974)}]{buenker1}%
  \BibitemOpen
  \bibfield  {author} {\bibinfo {author} {\bibfnamefont {R.~J.}\ \bibnamefont
  {Buenker}}\ and\ \bibinfo {author} {\bibfnamefont {S.~D.}\ \bibnamefont
  {Peyerimholf}},\ }\href@noop {} {\bibfield  {journal} {\bibinfo  {journal}
  {Theor. Chim. Acta}\ }\textbf {\bibinfo {volume} {35}},\ \bibinfo {pages}
  {33} (\bibinfo {year} {1974})}\BibitemShut {NoStop}%
\bibitem [{\citenamefont {Buenker}, \citenamefont {Peyerimholf},\ and\
  \citenamefont {Bruna}(1981)}]{buenker-book}%
  \BibitemOpen
  \bibfield  {author} {\bibinfo {author} {\bibfnamefont {R.~J.}\ \bibnamefont
  {Buenker}}, \bibinfo {author} {\bibfnamefont {S.~D.}\ \bibnamefont
  {Peyerimholf}}, \ and\ \bibinfo {author} {\bibfnamefont {P.~J.}\ \bibnamefont
  {Bruna}},\ }\href@noop {} {\emph {\bibinfo {title} {Comp. Theor. Org.
  Chem.}}}\ (\bibinfo  {publisher} {Reidel},\ \bibinfo {address} {Dordrecht},\
  \bibinfo {year} {1981})\ p.~\bibinfo {pages} {55}\BibitemShut {NoStop}%
\bibitem [{\citenamefont {Evangelisti}, \citenamefont {Daudey},\ and\
  \citenamefont {Malrieu}(1983)}]{three_class_CIPSI}%
  \BibitemOpen
  \bibfield  {author} {\bibinfo {author} {\bibfnamefont {S.}~\bibnamefont
  {Evangelisti}}, \bibinfo {author} {\bibfnamefont {J.-P.}\ \bibnamefont
  {Daudey}}, \ and\ \bibinfo {author} {\bibfnamefont {J.-P.}\ \bibnamefont
  {Malrieu}},\ }\href {\doibase https://doi.org/10.1016/0301-0104(83)85011-3}
  {\bibfield  {journal} {\bibinfo  {journal} {Chem. Phys.}\ }\textbf {\bibinfo
  {volume} {75}},\ \bibinfo {pages} {91 } (\bibinfo {year} {1983})}\BibitemShut
  {NoStop}%
\bibitem [{\citenamefont {Harrison}(1991)}]{harrison}%
  \BibitemOpen
  \bibfield  {author} {\bibinfo {author} {\bibfnamefont {R.~J.}\ \bibnamefont
  {Harrison}},\ }\href@noop {} {\bibfield  {journal} {\bibinfo  {journal} {J.
  Chem. Phys.}\ }\textbf {\bibinfo {volume} {94}},\ \bibinfo {pages} {5021}
  (\bibinfo {year} {1991})}\BibitemShut {NoStop}%
\bibitem [{\citenamefont {Holmes}, \citenamefont {Tubman},\ and\ \citenamefont
  {Umrigar}(2016)}]{hbci}%
  \BibitemOpen
  \bibfield  {author} {\bibinfo {author} {\bibfnamefont {A.~A.}\ \bibnamefont
  {Holmes}}, \bibinfo {author} {\bibfnamefont {N.~M.}\ \bibnamefont {Tubman}},
  \ and\ \bibinfo {author} {\bibfnamefont {C.~J.}\ \bibnamefont {Umrigar}},\
  }\href {\doibase 10.1021/acs.jctc.6b00407} {\bibfield  {journal} {\bibinfo
  {journal} {J. Chem. Theory Comput.}\ }\textbf {\bibinfo {volume} {12}},\
  \bibinfo {pages} {3674} (\bibinfo {year} {2016})}\BibitemShut {NoStop}%
\bibitem [{\citenamefont {Kato}(1957)}]{kato}%
  \BibitemOpen
  \bibfield  {author} {\bibinfo {author} {\bibfnamefont {T.}~\bibnamefont
  {Kato}},\ }\href@noop {} {\bibfield  {journal} {\bibinfo  {journal} {Comm.
  Pure Appl. Math.}\ }\textbf {\bibinfo {volume} {10}},\ \bibinfo {pages} {151}
  (\bibinfo {year} {1957})}\BibitemShut {NoStop}%
\bibitem [{\citenamefont {Hylleraas}(1929)}]{hylleraas}%
  \BibitemOpen
  \bibfield  {author} {\bibinfo {author} {\bibfnamefont {E.}~\bibnamefont
  {Hylleraas}},\ }\href@noop {} {\bibfield  {journal} {\bibinfo  {journal} {Z.
  Physik}\ }\textbf {\bibinfo {volume} {54}},\ \bibinfo {pages} {347} (\bibinfo
  {year} {1929})}\BibitemShut {NoStop}%
\bibitem [{\citenamefont {Hättig}\ \emph {et~al.}(2012)\citenamefont
  {Hättig}, \citenamefont {Klopper}, \citenamefont {Köhn},\ and\
  \citenamefont {Tew}}]{rev_f12_tew}%
  \BibitemOpen
  \bibfield  {author} {\bibinfo {author} {\bibfnamefont {C.}~\bibnamefont
  {Hättig}}, \bibinfo {author} {\bibfnamefont {W.}~\bibnamefont {Klopper}},
  \bibinfo {author} {\bibfnamefont {A.}~\bibnamefont {Köhn}}, \ and\ \bibinfo
  {author} {\bibfnamefont {D.~P.}\ \bibnamefont {Tew}},\ }\href {\doibase
  10.1021/cr200168z} {\bibfield  {journal} {\bibinfo  {journal} {Chem. Rev.}\
  }\textbf {\bibinfo {volume} {112}},\ \bibinfo {pages} {4} (\bibinfo {year}
  {2012})}\BibitemShut {NoStop}%
\bibitem [{\citenamefont {Kong}, \citenamefont {Bischoff},\ and\ \citenamefont
  {Valeev}(2012)}]{rev_f12_vallev}%
  \BibitemOpen
  \bibfield  {author} {\bibinfo {author} {\bibfnamefont {L.}~\bibnamefont
  {Kong}}, \bibinfo {author} {\bibfnamefont {F.~A.}\ \bibnamefont {Bischoff}},
  \ and\ \bibinfo {author} {\bibfnamefont {E.~F.}\ \bibnamefont {Valeev}},\
  }\href {\doibase 10.1021/cr200204r} {\bibfield  {journal} {\bibinfo
  {journal} {Chem. Rev.}\ }\textbf {\bibinfo {volume} {112}},\ \bibinfo {pages}
  {75} (\bibinfo {year} {2012})}\BibitemShut {NoStop}%
\bibitem [{\citenamefont {Grüneis}\ \emph {et~al.}(2017)\citenamefont
  {Grüneis}, \citenamefont {Hirata}, \citenamefont {Ohnishi},\ and\
  \citenamefont {Ten-no}}]{rev_f12_gruneis}%
  \BibitemOpen
  \bibfield  {author} {\bibinfo {author} {\bibfnamefont {A.}~\bibnamefont
  {Grüneis}}, \bibinfo {author} {\bibfnamefont {S.}~\bibnamefont {Hirata}},
  \bibinfo {author} {\bibfnamefont {Y.-Y.}\ \bibnamefont {Ohnishi}}, \ and\
  \bibinfo {author} {\bibfnamefont {S.}~\bibnamefont {Ten-no}},\ }\href
  {\doibase 10.1063/1.4976974} {\bibfield  {journal} {\bibinfo  {journal} {J.
  Chem. Phys.}\ }\textbf {\bibinfo {volume} {146}},\ \bibinfo {pages} {080901}
  (\bibinfo {year} {2017})}\BibitemShut {NoStop}%
\bibitem [{\citenamefont {Hohenberg}\ and\ \citenamefont
  {Kohn}(1964)}]{hk_theorem}%
  \BibitemOpen
  \bibfield  {author} {\bibinfo {author} {\bibfnamefont {P.}~\bibnamefont
  {Hohenberg}}\ and\ \bibinfo {author} {\bibfnamefont {W.}~\bibnamefont
  {Kohn}},\ }\href {\doibase 10.1103/PhysRev.136.B864} {\bibfield  {journal}
  {\bibinfo  {journal} {Phys. Rev.}\ }\textbf {\bibinfo {volume} {136}},\
  \bibinfo {pages} {B864} (\bibinfo {year} {1964})}\BibitemShut {NoStop}%
\bibitem [{\citenamefont {Kohn}\ and\ \citenamefont {Sham}(1965)}]{ks_dft}%
  \BibitemOpen
  \bibfield  {author} {\bibinfo {author} {\bibfnamefont {W.}~\bibnamefont
  {Kohn}}\ and\ \bibinfo {author} {\bibfnamefont {L.~J.}\ \bibnamefont
  {Sham}},\ }\href {\doibase 10.1103/PhysRev.140.A1133} {\bibfield  {journal}
  {\bibinfo  {journal} {Phys. Rev.}\ }\textbf {\bibinfo {volume} {140}},\
  \bibinfo {pages} {A1133} (\bibinfo {year} {1965})}\BibitemShut {NoStop}%
\bibitem [{\citenamefont {Becke}(1993)}]{Bec-JCP-93a}%
  \BibitemOpen
  \bibfield  {author} {\bibinfo {author} {\bibfnamefont {A.~D.}\ \bibnamefont
  {Becke}},\ }\href@noop {} {\bibfield  {journal} {\bibinfo  {journal} {J.
  Chem. Phys.}\ }\textbf {\bibinfo {volume} {98}},\ \bibinfo {pages} {1372}
  (\bibinfo {year} {1993})}\BibitemShut {NoStop}%
\bibitem [{\citenamefont {Goerigk}\ and\ \citenamefont
  {Grimme}(2014)}]{GoeGri-WIRE-14}%
  \BibitemOpen
  \bibfield  {author} {\bibinfo {author} {\bibfnamefont {L.}~\bibnamefont
  {Goerigk}}\ and\ \bibinfo {author} {\bibfnamefont {S.}~\bibnamefont
  {Grimme}},\ }\href@noop {} {\bibfield  {journal} {\bibinfo  {journal} {WIREs
  Comput. Mol. Sci.}\ }\textbf {\bibinfo {volume} {4}},\ \bibinfo {pages} {576}
  (\bibinfo {year} {2014})}\BibitemShut {NoStop}%
\bibitem [{\citenamefont {Jones}\ and\ \citenamefont
  {Gunnarsson}(1989)}]{gunnarson_dft_rev}%
  \BibitemOpen
  \bibfield  {author} {\bibinfo {author} {\bibfnamefont {R.~O.}\ \bibnamefont
  {Jones}}\ and\ \bibinfo {author} {\bibfnamefont {O.}~\bibnamefont
  {Gunnarsson}},\ }\href {\doibase 10.1103/RevModPhys.61.689} {\bibfield
  {journal} {\bibinfo  {journal} {Rev. Mod. Phys.}\ }\textbf {\bibinfo {volume}
  {61}},\ \bibinfo {pages} {689} (\bibinfo {year} {1989})}\BibitemShut
  {NoStop}%
\bibitem [{\citenamefont {Toulouse}, \citenamefont {Colonna},\ and\
  \citenamefont {Savin}(2004)}]{rs_dft_toul_colo_savin}%
  \BibitemOpen
  \bibfield  {author} {\bibinfo {author} {\bibfnamefont {J.}~\bibnamefont
  {Toulouse}}, \bibinfo {author} {\bibfnamefont {F.}~\bibnamefont {Colonna}}, \
  and\ \bibinfo {author} {\bibfnamefont {A.}~\bibnamefont {Savin}},\ }\href
  {\doibase 10.1103/PhysRevA.70.062505} {\bibfield  {journal} {\bibinfo
  {journal} {Phys. Rev. A}\ }\textbf {\bibinfo {volume} {70}},\ \bibinfo
  {pages} {062505} (\bibinfo {year} {2004})}\BibitemShut {NoStop}%
\bibitem [{\citenamefont {Franck}\ \emph {et~al.}(2015)\citenamefont {Franck},
  \citenamefont {Mussard}, \citenamefont {Luppi},\ and\ \citenamefont
  {Toulouse}}]{basis_set_rs_dft}%
  \BibitemOpen
  \bibfield  {author} {\bibinfo {author} {\bibfnamefont {O.}~\bibnamefont
  {Franck}}, \bibinfo {author} {\bibfnamefont {B.}~\bibnamefont {Mussard}},
  \bibinfo {author} {\bibfnamefont {E.}~\bibnamefont {Luppi}}, \ and\ \bibinfo
  {author} {\bibfnamefont {J.}~\bibnamefont {Toulouse}},\ }\href {\doibase
  10.1063/1.4907920} {\bibfield  {journal} {\bibinfo  {journal} {J. Chem.
  Phys.}\ }\textbf {\bibinfo {volume} {142}},\ \bibinfo {pages} {074107}
  (\bibinfo {year} {2015})}\BibitemShut {NoStop}%
\bibitem [{\citenamefont {\'Angy\'an}\ \emph {et~al.}(2005)\citenamefont
  {\'Angy\'an}, \citenamefont {Gerber}, \citenamefont {Savin},\ and\
  \citenamefont {Toulouse}}]{AngGerSavTou-PRA-05}%
  \BibitemOpen
  \bibfield  {author} {\bibinfo {author} {\bibfnamefont {J.~G.}\ \bibnamefont
  {\'Angy\'an}}, \bibinfo {author} {\bibfnamefont {I.~C.}\ \bibnamefont
  {Gerber}}, \bibinfo {author} {\bibfnamefont {A.}~\bibnamefont {Savin}}, \
  and\ \bibinfo {author} {\bibfnamefont {J.}~\bibnamefont {Toulouse}},\
  }\href@noop {} {\bibfield  {journal} {\bibinfo  {journal} {Phys. Rev. A}\
  }\textbf {\bibinfo {volume} {72}},\ \bibinfo {pages} {012510} (\bibinfo
  {year} {2005})}\BibitemShut {NoStop}%
\bibitem [{\citenamefont {Goll}, \citenamefont {Werner},\ and\ \citenamefont
  {Stoll}(2005)}]{GolWerSto-PCCP-05}%
  \BibitemOpen
  \bibfield  {author} {\bibinfo {author} {\bibfnamefont {E.}~\bibnamefont
  {Goll}}, \bibinfo {author} {\bibfnamefont {H.-J.}\ \bibnamefont {Werner}}, \
  and\ \bibinfo {author} {\bibfnamefont {H.}~\bibnamefont {Stoll}},\
  }\href@noop {} {\bibfield  {journal} {\bibinfo  {journal} {Phys. Chem. Chem.
  Phys.}\ }\textbf {\bibinfo {volume} {7}},\ \bibinfo {pages} {3917} (\bibinfo
  {year} {2005})}\BibitemShut {NoStop}%
\bibitem [{\citenamefont {Toulouse}\ \emph {et~al.}(2009)\citenamefont
  {Toulouse}, \citenamefont {Gerber}, \citenamefont {Jansen}, \citenamefont
  {Savin},\ and\ \citenamefont {\'Angy\'an}}]{TouGerJanSavAng-PRL-09}%
  \BibitemOpen
  \bibfield  {author} {\bibinfo {author} {\bibfnamefont {J.}~\bibnamefont
  {Toulouse}}, \bibinfo {author} {\bibfnamefont {I.~C.}\ \bibnamefont
  {Gerber}}, \bibinfo {author} {\bibfnamefont {G.}~\bibnamefont {Jansen}},
  \bibinfo {author} {\bibfnamefont {A.}~\bibnamefont {Savin}}, \ and\ \bibinfo
  {author} {\bibfnamefont {J.~G.}\ \bibnamefont {\'Angy\'an}},\ }\href@noop {}
  {\bibfield  {journal} {\bibinfo  {journal} {Phys. Rev. Lett.}\ }\textbf
  {\bibinfo {volume} {102}},\ \bibinfo {pages} {096404} (\bibinfo {year}
  {2009})}\BibitemShut {NoStop}%
\bibitem [{\citenamefont {Janesko}, \citenamefont {Henderson},\ and\
  \citenamefont {Scuseria}(2009)}]{JanHenScu-JCP-09}%
  \BibitemOpen
  \bibfield  {author} {\bibinfo {author} {\bibfnamefont {B.~G.}\ \bibnamefont
  {Janesko}}, \bibinfo {author} {\bibfnamefont {T.~M.}\ \bibnamefont
  {Henderson}}, \ and\ \bibinfo {author} {\bibfnamefont {G.~E.}\ \bibnamefont
  {Scuseria}},\ }\href@noop {} {\bibfield  {journal} {\bibinfo  {journal} {J.
  Chem. Phys.}\ }\textbf {\bibinfo {volume} {130}},\ \bibinfo {pages} {081105}
  (\bibinfo {year} {2009})}\BibitemShut {NoStop}%
\bibitem [{\citenamefont {Leininger}\ \emph {et~al.}(1997)\citenamefont
  {Leininger}, \citenamefont {Stoll}, \citenamefont {Werner},\ and\
  \citenamefont {Savin}}]{LeiStoWerSav-CPL-97}%
  \BibitemOpen
  \bibfield  {author} {\bibinfo {author} {\bibfnamefont {T.}~\bibnamefont
  {Leininger}}, \bibinfo {author} {\bibfnamefont {H.}~\bibnamefont {Stoll}},
  \bibinfo {author} {\bibfnamefont {H.-J.}\ \bibnamefont {Werner}}, \ and\
  \bibinfo {author} {\bibfnamefont {A.}~\bibnamefont {Savin}},\ }\href@noop {}
  {\bibfield  {journal} {\bibinfo  {journal} {Chem. Phys. Lett.}\ }\textbf
  {\bibinfo {volume} {{275}}},\ \bibinfo {pages} {151} (\bibinfo {year}
  {1997})}\BibitemShut {NoStop}%
\bibitem [{\citenamefont {Fromager}, \citenamefont {Toulouse},\ and\
  \citenamefont {Jensen}(2007)}]{FroTouJen-JCP-07}%
  \BibitemOpen
  \bibfield  {author} {\bibinfo {author} {\bibfnamefont {E.}~\bibnamefont
  {Fromager}}, \bibinfo {author} {\bibfnamefont {J.}~\bibnamefont {Toulouse}},
  \ and\ \bibinfo {author} {\bibfnamefont {H.~J.~A.}\ \bibnamefont {Jensen}},\
  }\href@noop {} {\bibfield  {journal} {\bibinfo  {journal} {J. Chem. Phys.}\
  }\textbf {\bibinfo {volume} {126}},\ \bibinfo {pages} {074111} (\bibinfo
  {year} {2007})}\BibitemShut {NoStop}%
\bibitem [{\citenamefont {Fromager}, \citenamefont {Cimiraglia},\ and\
  \citenamefont {Jensen}(2010)}]{fromager_rs_nevpt2}%
  \BibitemOpen
  \bibfield  {author} {\bibinfo {author} {\bibfnamefont {E.}~\bibnamefont
  {Fromager}}, \bibinfo {author} {\bibfnamefont {R.}~\bibnamefont
  {Cimiraglia}}, \ and\ \bibinfo {author} {\bibfnamefont {H.~J.~A.}\
  \bibnamefont {Jensen}},\ }\href {\doibase 10.1103/PhysRevA.81.024502}
  {\bibfield  {journal} {\bibinfo  {journal} {Phys. Rev. A}\ }\textbf {\bibinfo
  {volume} {81}},\ \bibinfo {pages} {024502} (\bibinfo {year}
  {2010})}\BibitemShut {NoStop}%
\bibitem [{\citenamefont {Hedegård}\ \emph {et~al.}(2015)\citenamefont
  {Hedegård}, \citenamefont {Knecht}, \citenamefont {Kielberg}, \citenamefont
  {Jensen},\ and\ \citenamefont {Reiher}}]{dmrg_rs_dft_1}%
  \BibitemOpen
  \bibfield  {author} {\bibinfo {author} {\bibfnamefont {E.~D.}\ \bibnamefont
  {Hedegård}}, \bibinfo {author} {\bibfnamefont {S.}~\bibnamefont {Knecht}},
  \bibinfo {author} {\bibfnamefont {J.~S.}\ \bibnamefont {Kielberg}}, \bibinfo
  {author} {\bibfnamefont {H.~J.~A.}\ \bibnamefont {Jensen}}, \ and\ \bibinfo
  {author} {\bibfnamefont {M.}~\bibnamefont {Reiher}},\ }\href {\doibase
  10.1063/1.4922295} {\bibfield  {journal} {\bibinfo  {journal} {J. Chem.
  Phys.}\ }\textbf {\bibinfo {volume} {142}},\ \bibinfo {pages} {224108}
  (\bibinfo {year} {2015})}\BibitemShut {NoStop}%
\bibitem [{\citenamefont {Kronik}\ \emph {et~al.}(2012)\citenamefont {Kronik},
  \citenamefont {Stein}, \citenamefont {Refaely-Abramson},\ and\ \citenamefont
  {Baer}}]{optimal_tuned_rsdft}%
  \BibitemOpen
  \bibfield  {author} {\bibinfo {author} {\bibfnamefont {L.}~\bibnamefont
  {Kronik}}, \bibinfo {author} {\bibfnamefont {T.}~\bibnamefont {Stein}},
  \bibinfo {author} {\bibfnamefont {S.}~\bibnamefont {Refaely-Abramson}}, \
  and\ \bibinfo {author} {\bibfnamefont {R.}~\bibnamefont {Baer}},\ }\href
  {\doibase 10.1021/ct2009363} {\bibfield  {journal} {\bibinfo  {journal} {J.
  Chem. Theory Comput.}\ }\textbf {\bibinfo {volume} {8}},\ \bibinfo {pages}
  {1515} (\bibinfo {year} {2012})}\BibitemShut {NoStop}%
\bibitem [{\citenamefont {Krukau}\ \emph {et~al.}(2008)\citenamefont {Krukau},
  \citenamefont {Scuseria}, \citenamefont {Perdew},\ and\ \citenamefont
  {Savin}}]{local_mu_hybrid_1}%
  \BibitemOpen
  \bibfield  {author} {\bibinfo {author} {\bibfnamefont {A.~V.}\ \bibnamefont
  {Krukau}}, \bibinfo {author} {\bibfnamefont {G.~E.}\ \bibnamefont
  {Scuseria}}, \bibinfo {author} {\bibfnamefont {J.~P.}\ \bibnamefont
  {Perdew}}, \ and\ \bibinfo {author} {\bibfnamefont {A.}~\bibnamefont
  {Savin}},\ }\href {\doibase 10.1063/1.2978377} {\bibfield  {journal}
  {\bibinfo  {journal} {J. Chem. Phys.}\ }\textbf {\bibinfo {volume} {129}},\
  \bibinfo {pages} {124103} (\bibinfo {year} {2008})}\BibitemShut {NoStop}%
\bibitem [{\citenamefont {Henderson}\ \emph {et~al.}(2009)\citenamefont
  {Henderson}, \citenamefont {Janesko}, \citenamefont {Scuseria},\ and\
  \citenamefont {Savin}}]{local_mu_hybrid_2}%
  \BibitemOpen
  \bibfield  {author} {\bibinfo {author} {\bibfnamefont {T.~M.}\ \bibnamefont
  {Henderson}}, \bibinfo {author} {\bibfnamefont {B.~G.}\ \bibnamefont
  {Janesko}}, \bibinfo {author} {\bibfnamefont {G.~E.}\ \bibnamefont
  {Scuseria}}, \ and\ \bibinfo {author} {\bibfnamefont {A.}~\bibnamefont
  {Savin}},\ }\href {\doibase 10.1002/qua.22049} {\bibfield  {journal}
  {\bibinfo  {journal} {Int. J. Quantum Chem.}\ }\textbf {\bibinfo {volume}
  {109}},\ \bibinfo {pages} {2023} (\bibinfo {year} {2009})}\BibitemShut
  {NoStop}%
\bibitem [{\citenamefont {Toulouse}, \citenamefont {Gori-Giorgi},\ and\
  \citenamefont {Savin}(2005)}]{Toulouse2005_ecmd}%
  \BibitemOpen
  \bibfield  {author} {\bibinfo {author} {\bibfnamefont {J.}~\bibnamefont
  {Toulouse}}, \bibinfo {author} {\bibfnamefont {P.}~\bibnamefont
  {Gori-Giorgi}}, \ and\ \bibinfo {author} {\bibfnamefont {A.}~\bibnamefont
  {Savin}},\ }\href {\doibase 10.1007/s00214-005-0688-2} {\bibfield  {journal}
  {\bibinfo  {journal} {Theor. Chem. Acc.}\ }\textbf {\bibinfo {volume}
  {114}},\ \bibinfo {pages} {305} (\bibinfo {year} {2005})}\BibitemShut
  {NoStop}%
\bibitem [{\citenamefont {Slater}(1951)}]{slater_exch_potential}%
  \BibitemOpen
  \bibfield  {author} {\bibinfo {author} {\bibfnamefont {J.~C.}\ \bibnamefont
  {Slater}},\ }\href {\doibase 10.1103/PhysRev.81.385} {\bibfield  {journal}
  {\bibinfo  {journal} {Phys. Rev.}\ }\textbf {\bibinfo {volume} {81}},\
  \bibinfo {pages} {385} (\bibinfo {year} {1951})}\BibitemShut {NoStop}%
\bibitem [{\citenamefont {Paziani}\ \emph {et~al.}(2006)\citenamefont
  {Paziani}, \citenamefont {Moroni}, \citenamefont {Gori-Giorgi},\ and\
  \citenamefont {Bachelet}}]{PazMorGorBac-PRB-06}%
  \BibitemOpen
  \bibfield  {author} {\bibinfo {author} {\bibfnamefont {S.}~\bibnamefont
  {Paziani}}, \bibinfo {author} {\bibfnamefont {S.}~\bibnamefont {Moroni}},
  \bibinfo {author} {\bibfnamefont {P.}~\bibnamefont {Gori-Giorgi}}, \ and\
  \bibinfo {author} {\bibfnamefont {G.~B.}\ \bibnamefont {Bachelet}},\
  }\href@noop {} {\bibfield  {journal} {\bibinfo  {journal} {Phys. Rev. B}\
  }\textbf {\bibinfo {volume} {73}},\ \bibinfo {pages} {155111} (\bibinfo
  {year} {2006})}\BibitemShut {NoStop}%
\bibitem [{\citenamefont {Rubio}, \citenamefont {Novoa},\ and\ \citenamefont
  {Illas}(1986)}]{Rubio198698}%
  \BibitemOpen
  \bibfield  {author} {\bibinfo {author} {\bibfnamefont {J.}~\bibnamefont
  {Rubio}}, \bibinfo {author} {\bibfnamefont {J.}~\bibnamefont {Novoa}}, \ and\
  \bibinfo {author} {\bibfnamefont {F.}~\bibnamefont {Illas}},\ }\href
  {\doibase http://dx.doi.org/10.1016/0009-2614(86)85123-5} {\bibfield
  {journal} {\bibinfo  {journal} {Chem. Phys. Lett.}\ }\textbf {\bibinfo
  {volume} {126}},\ \bibinfo {pages} {98 } (\bibinfo {year}
  {1986})}\BibitemShut {NoStop}%
\bibitem [{\citenamefont {Cimiraglia}\ and\ \citenamefont
  {Persico}(1987)}]{cimiraglia_cipsi}%
  \BibitemOpen
  \bibfield  {author} {\bibinfo {author} {\bibfnamefont {R.}~\bibnamefont
  {Cimiraglia}}\ and\ \bibinfo {author} {\bibfnamefont {M.}~\bibnamefont
  {Persico}},\ }\href {\doibase 10.1002/jcc.540080105} {\bibfield  {journal}
  {\bibinfo  {journal} {J. Comp. Chem.}\ }\textbf {\bibinfo {volume} {8}},\
  \bibinfo {pages} {39} (\bibinfo {year} {1987})}\BibitemShut {NoStop}%
\bibitem [{\citenamefont {Angeli}\ and\ \citenamefont
  {Persico}(1997)}]{cele_cipsi_zeroth_order}%
  \BibitemOpen
  \bibfield  {author} {\bibinfo {author} {\bibfnamefont {C.}~\bibnamefont
  {Angeli}}\ and\ \bibinfo {author} {\bibfnamefont {M.}~\bibnamefont
  {Persico}},\ }\href {\doibase 10.1007/s002140050285} {\bibfield  {journal}
  {\bibinfo  {journal} {Theor. Chem. Acc.}\ }\textbf {\bibinfo {volume} {98}},\
  \bibinfo {pages} {117} (\bibinfo {year} {1997})}\BibitemShut {NoStop}%
\bibitem [{\citenamefont {Angeli}, \citenamefont {Cimiraglia},\ and\
  \citenamefont {Malrieu}(2000)}]{Angeli2000472}%
  \BibitemOpen
  \bibfield  {author} {\bibinfo {author} {\bibfnamefont {C.}~\bibnamefont
  {Angeli}}, \bibinfo {author} {\bibfnamefont {R.}~\bibnamefont {Cimiraglia}},
  \ and\ \bibinfo {author} {\bibfnamefont {J.-P.}\ \bibnamefont {Malrieu}},\
  }\href {\doibase http://dx.doi.org/10.1016/S0009-2614(99)01458-X} {\bibfield
  {journal} {\bibinfo  {journal} {Chem. Phys. Lett.}\ }\textbf {\bibinfo
  {volume} {317}},\ \bibinfo {pages} {472 } (\bibinfo {year}
  {2000})}\BibitemShut {NoStop}%
\bibitem [{\citenamefont {Giner}, \citenamefont {Scemama},\ and\ \citenamefont
  {Caffarel}(2013)}]{canadian}%
  \BibitemOpen
  \bibfield  {author} {\bibinfo {author} {\bibfnamefont {E.}~\bibnamefont
  {Giner}}, \bibinfo {author} {\bibfnamefont {A.}~\bibnamefont {Scemama}}, \
  and\ \bibinfo {author} {\bibfnamefont {M.}~\bibnamefont {Caffarel}},\
  }\href@noop {} {\bibfield  {journal} {\bibinfo  {journal} {Can. J. Chem.}\
  }\textbf {\bibinfo {volume} {91}},\ \bibinfo {pages} {879} (\bibinfo {year}
  {2013})}\BibitemShut {NoStop}%
\bibitem [{\citenamefont {Scemama}\ \emph {et~al.}(2014)\citenamefont
  {Scemama}, \citenamefont {Applencourt}, \citenamefont {Giner},\ and\
  \citenamefont {Caffarel}}]{atoms_3d}%
  \BibitemOpen
  \bibfield  {author} {\bibinfo {author} {\bibfnamefont {A.}~\bibnamefont
  {Scemama}}, \bibinfo {author} {\bibfnamefont {T.}~\bibnamefont
  {Applencourt}}, \bibinfo {author} {\bibfnamefont {E.}~\bibnamefont {Giner}},
  \ and\ \bibinfo {author} {\bibfnamefont {M.}~\bibnamefont {Caffarel}},\
  }\href@noop {} {\bibfield  {journal} {\bibinfo  {journal} {J. Chem. Phys.}\
  }\textbf {\bibinfo {volume} {141}},\ \bibinfo {eid} {244110} (\bibinfo {year}
  {2014})}\BibitemShut {NoStop}%
\bibitem [{\citenamefont {Giner}, \citenamefont {Scemama},\ and\ \citenamefont
  {Caffarel}(2015)}]{f2_dmc}%
  \BibitemOpen
  \bibfield  {author} {\bibinfo {author} {\bibfnamefont {E.}~\bibnamefont
  {Giner}}, \bibinfo {author} {\bibfnamefont {A.}~\bibnamefont {Scemama}}, \
  and\ \bibinfo {author} {\bibfnamefont {M.}~\bibnamefont {Caffarel}},\
  }\href@noop {} {\bibfield  {journal} {\bibinfo  {journal} {J. Chem. Phys.}\
  }\textbf {\bibinfo {volume} {142}},\ \bibinfo {eid} {044115} (\bibinfo {year}
  {2015})}\BibitemShut {NoStop}%
\bibitem [{\citenamefont {Giner}, \citenamefont {Assaraf},\ and\ \citenamefont
  {Toulouse}(2016)}]{atoms_dmc_julien}%
  \BibitemOpen
  \bibfield  {author} {\bibinfo {author} {\bibfnamefont {E.}~\bibnamefont
  {Giner}}, \bibinfo {author} {\bibfnamefont {R.}~\bibnamefont {Assaraf}}, \
  and\ \bibinfo {author} {\bibfnamefont {J.}~\bibnamefont {Toulouse}},\ }\href
  {\doibase 10.1080/00268976.2016.1149630} {\bibfield  {journal} {\bibinfo
  {journal} {Mol. Phys.}\ }\textbf {\bibinfo {volume} {114}},\ \bibinfo {pages}
  {910} (\bibinfo {year} {2016})}\BibitemShut {NoStop}%
\bibitem [{\citenamefont {Epstein}(1926)}]{epstein}%
  \BibitemOpen
  \bibfield  {author} {\bibinfo {author} {\bibfnamefont {P.~S.}\ \bibnamefont
  {Epstein}},\ }\href@noop {} {\bibfield  {journal} {\bibinfo  {journal} {Phys.
  Rev.}\ }\textbf {\bibinfo {volume} {28}},\ \bibinfo {pages} {695} (\bibinfo
  {year} {1926})}\BibitemShut {NoStop}%
\bibitem [{\citenamefont {Nesbet}(1955)}]{nesbet}%
  \BibitemOpen
  \bibfield  {author} {\bibinfo {author} {\bibfnamefont {R.~K.}\ \bibnamefont
  {Nesbet}},\ }\href@noop {} {\bibfield  {journal} {\bibinfo  {journal} {Proc.
  R. Soc. A}\ }\textbf {\bibinfo {volume} {230}},\ \bibinfo {pages} {312}
  (\bibinfo {year} {1955})}\BibitemShut {NoStop}%
\bibitem [{\citenamefont {Garniron}\ \emph {et~al.}(2017)\citenamefont
  {Garniron}, \citenamefont {Scemama}, \citenamefont {Loos},\ and\
  \citenamefont {Caffarel}}]{stochastic_pt_yan}%
  \BibitemOpen
  \bibfield  {author} {\bibinfo {author} {\bibfnamefont {Y.}~\bibnamefont
  {Garniron}}, \bibinfo {author} {\bibfnamefont {A.}~\bibnamefont {Scemama}},
  \bibinfo {author} {\bibfnamefont {P.-F.}\ \bibnamefont {Loos}}, \ and\
  \bibinfo {author} {\bibfnamefont {M.}~\bibnamefont {Caffarel}},\ }\href
  {\doibase 10.1063/1.4992127} {\bibfield  {journal} {\bibinfo  {journal} {J.
  Chem. Phys.}\ }\textbf {\bibinfo {volume} {147}},\ \bibinfo {pages} {034101}
  (\bibinfo {year} {2017})}\BibitemShut {NoStop}%
\bibitem [{\citenamefont {Scemama}\ \emph {et~al.}(2016)\citenamefont
  {Scemama}, \citenamefont {Applencourt}, \citenamefont {Garniron},
  \citenamefont {Giner}, \citenamefont {David},\ and\ \citenamefont
  {Caffarel}}]{qp}%
  \BibitemOpen
  \bibfield  {author} {\bibinfo {author} {\bibfnamefont {A.}~\bibnamefont
  {Scemama}}, \bibinfo {author} {\bibfnamefont {T.}~\bibnamefont
  {Applencourt}}, \bibinfo {author} {\bibfnamefont {Y.}~\bibnamefont
  {Garniron}}, \bibinfo {author} {\bibfnamefont {E.}~\bibnamefont {Giner}},
  \bibinfo {author} {\bibfnamefont {G.}~\bibnamefont {David}}, \ and\ \bibinfo
  {author} {\bibfnamefont {M.}~\bibnamefont {Caffarel}},\ }\href {\doibase
  https://github.com/LCPQ/quantum_package} {\enquote {\bibinfo {title} {Quantum
  package v1.0},}\ } (\bibinfo {year} {2016})\BibitemShut {NoStop}%
\bibitem [{\citenamefont {Booth}\ and\ \citenamefont {Alavi}(2010)}]{ip_ali}%
  \BibitemOpen
  \bibfield  {author} {\bibinfo {author} {\bibfnamefont {G.~H.}\ \bibnamefont
  {Booth}}\ and\ \bibinfo {author} {\bibfnamefont {A.}~\bibnamefont {Alavi}},\
  }\href {\doibase 10.1063/1.3407895} {\bibfield  {journal} {\bibinfo
  {journal} {J. Chem. Phys.}\ }\textbf {\bibinfo {volume} {132}},\ \bibinfo
  {pages} {174104} (\bibinfo {year} {2010})}\BibitemShut {NoStop}%
\bibitem [{\citenamefont {Chakravorty}\ \emph {et~al.}(1993)\citenamefont
  {Chakravorty}, \citenamefont {Gwaltney}, \citenamefont {Davidson},
  \citenamefont {Parpia},\ and\ \citenamefont {Froese~Fischer}}]{exact_atoms}%
  \BibitemOpen
  \bibfield  {author} {\bibinfo {author} {\bibfnamefont {S.~J.}\ \bibnamefont
  {Chakravorty}}, \bibinfo {author} {\bibfnamefont {S.~R.}\ \bibnamefont
  {Gwaltney}}, \bibinfo {author} {\bibfnamefont {E.~R.}\ \bibnamefont
  {Davidson}}, \bibinfo {author} {\bibfnamefont {F.~A.}\ \bibnamefont
  {Parpia}}, \ and\ \bibinfo {author} {\bibfnamefont {C.}~\bibnamefont
  {Froese~Fischer}},\ }\href {\doibase 10.1103/PhysRevA.47.3649} {\bibfield
  {journal} {\bibinfo  {journal} {Phys. Rev. A}\ }\textbf {\bibinfo {volume}
  {47}},\ \bibinfo {pages} {3649} (\bibinfo {year} {1993})}\BibitemShut
  {NoStop}%
\end{thebibliography}
 \end{document}